\documentclass[11pt,a4paper]{article}
\usepackage[T1]{fontenc}
\usepackage[utf8]{inputenc}
\usepackage{lmodern}
\usepackage[margin=1.06in]{geometry}
\usepackage{microtype}
\usepackage{amsmath,amssymb,mathtools,amsthm}
\usepackage{graphicx,booktabs,tabularx,array}
\usepackage[round,authoryear]{natbib}
\usepackage{caption}
\usepackage[section]{placeins}
\usepackage{fancyhdr}
\usepackage{xcolor}
\usepackage{url}
\usepackage[colorlinks=true,linkcolor=blue!40!black,citecolor=blue!40!black,urlcolor=blue!40!black]{hyperref}
\graphicspath{{figures/}}
\newcommand{\calH}{\mathcal H}

\newcommand{\E}{\mathbb E}
\newcommand{\ind}{\mathbf 1}
\newcommand{\dd}{\,\mathrm d}
\newcommand{\NN}{\mathcal N}
\DeclareMathOperator{\softplus}{softplus}

\DeclareMathOperator{\Var}{Var}

\DeclareMathOperator{\GammaD}{Gamma}
\title{\bfseries \huge Neural Marked Point Processes for Health Insurance Claims}
\author{Eva Flonner}
\date{November 2026}

\begin{document}
\maketitle
\vspace{-1.0em}
\begin{abstract}
Health insurance claims exhibit temporal dependence, as previous claims may provide information about the timing, type and amount of subsequent claims. Such dependencies are difficult to capture with traditional actuarial models based on aggregated claim counts. We investigate a neural marked point process in which a causal Transformer represents individual claim histories and determines type-specific conditional intensities, allowing for self-excitation and interactions between claim types. The framework also incorporates conditional severity distributions, reporting delays and stochastic medical-cost inflation. We develop a likelihood-based calibration procedure that accounts for incomplete claim reporting and evaluate the model using both simulated and empirical data. A simulation study with 10,000 insurance contracts examines the recovery of known claim dynamics, while an application to MEPS-derived medical-event data assesses predictive performance. The fitted model reproduces the main patterns in claim frequencies and severities, although some discrepancies remain in the tails of the distributions. On the MEPS data, the neural model achieves better predictive performance than two generalized linear model benchmarks in terms of claim-count deviance and expenditure discrimination.
\end{abstract}
\noindent\textbf{Keywords:} neural marked point process; claim histories; likelihood calibration; self-excitation; reporting delay; medical inflation; Medical Expenditure Panel Survey.

\section{Introduction}\label{sec:intro}

Consider two beneficiaries who each generate the same number of health insurance claims in a year. For one, several outpatient claims follow an inpatient admission within a short period; for the other, the claims are spread over several months. The annual totals are identical, but the claim histories differ in their timing and sequence. In classical actuarial frequency models, claims are commonly aggregated into annual counts, so these differences are not explicitly taken into account.

This paper investigates whether neural marked point processes provide a useful alternative to classical actuarial models for predicting health-insurance claims. In particular, we examine whether incorporating the timing, type and amount of previous claims improves predictive performance compared with generalized linear models, which typically rely on aggregated claim counts and policyholder characteristics. To this end, we consider a neural marked point process in which a causal Transformer summarizes individual claim histories and determines type-specific conditional intensities. This allows the model to capture self-excitation and more general dependencies between successive claims without imposing a particular functional form. The framework further includes conditional claim severity distributions, reporting delays and a stochastic medical-cost index. We evaluate the model through a simulation study with a known data-generating process and an empirical application to health-insurance claims data, focusing on calibration quality and out-of-sample predictive performance.

The modeling choice builds on studies of predictive learning from insurance claims, which show that previous utilization and its temporal structure can be informative for future expenditure \citep{morid2019,morid2020,dreweboss2022}. More closely related to our approach are marked point-process models of medical event histories and neural temporal point processes \citep{enguehard2020,mews2023}. Section~\ref{sec:literature} reviews these contributions and discusses their different modeling approaches and prediction targets. Our contribution lies in developing and evaluating a neural marked point-process framework tailored to the characteristics of health insurance claims. We investigate whether the fitted model adequately reproduces the distributions of claim occurrences and amounts at both the individual and aggregate levels, and whether incorporating individual claim histories improves predictive performance relative to conventional actuarial benchmarks.

Calibrating a neural marked point process requires accounting not only for the claims that occur, but also for the periods during which no claims are observed. This is particularly challenging in health insurance, where claims are often reported with a delay. At any given valuation date, some claims may already have occurred without being known to the insurer. Since the occurrence of a claim can influence the intensity of subsequent claims, these unreported events complicate the estimation of the model. In addition, changes in claim amounts over time may reflect medical inflation as well as changes in the characteristics of insured individuals and the types of claims they generate. We address these issues by incorporating reporting delays and a stochastic medical-cost index into the modeling framework, and by developing a likelihood-based calibration procedure that accounts for the available claim histories.

Two studies are used to evaluate calibration. The first is a controlled simulation of 10,000 health-insurance contracts generated from a process specified independently of the neural network. It contains self- and cross-excitation, individual heterogeneity, claim severity, delayed reporting and medical inflation. Comparing the fitted model with this known generator shows which features the method recovers and where it fails. The second study uses processed MEPS-based event histories. It assesses in-sample frequencies, between-person count heterogeneity and claim-amount distributions, followed by forecasts for a held-out calendar year. The forecast results are compared with two generalized linear models, including one with coarse summaries of prior experience. Because the datasets answer different questions, simulation recovery and observational predictive performance are reported separately throughout.

The mathematical specification and likelihood are developed first, followed by the calibration algorithm. We then present the controlled simulation and the MEPS-based results. The latter records are medical utilization and expenditure events, not verified insurer claim-receipt records; consequently, they cannot independently validate the reporting-delay model. The paper ends by discussing the remaining uncertainty about the benefits of neural history dependence, especially the need for direct excitation diagnostics and matched baselines.

\section{Related literature}\label{sec:literature}

Machine learning has been applied extensively to health-insurance records, but the statistical target in much of this work is an individual's total expenditure over a subsequent year rather than the sequence of events that generates it. In an early comparison based on US insurer records, \citet{morid2018} evaluated several supervised-learning approaches to annual healthcare cost prediction. Gradient boosting performed best overall in their study, while neural networks and ridge regression were competitive for high-cost individuals. Using French national claims data, \citet{vimont2022} likewise compared a generalized linear model, random forests and a neural network for individual expenditure prediction. They found that random forests performed best when richer sets of predictors were available. Together, these studies illustrate that more flexible models do not necessarily provide better predictions and that their performance must be assessed empirically. Accordingly, we evaluate the predictive performance of our neural marked point process against two generalized linear model benchmarks.

A separate strand of the literature examines the predictive value of temporal information in insurance claims. \citet{morid2019} used fine-grained time windows and features capturing changes in utilization and expenditure, demonstrating improvements in subsequent cost prediction compared with coarser representations. \citet{morid2020} employed convolutional neural networks to identify patterns in multivariate time series of medical and pharmacy claims. Both studies provide evidence that information about previous utilization can improve expenditure predictions. However, their approaches represent claim histories in discrete time and focus on predicting aggregate future costs. In contrast, we model individual claim occurrences through history-dependent, type-specific conditional intensities, together with conditional distributions for claim amounts. This allows the timing, type and severity of previous claims to inform predictions of subsequent claim occurrences and expenditures.

Further studies have investigated the prediction of healthcare expenditures and medical events using large administrative claims databases. \citet{dreweboss2022} trained a deep neural network on records from approximately 1.4 million members of German statutory health insurers and reported improvements in one-year expenditure prediction over ridge regression and other benchmark models. \citet{moridsheng2025} examined the role of patient heterogeneity and different sources of claims information in deep-learning-based expenditure prediction. While these studies focus on predicting future healthcare costs, \citet{yoshimoto2024} considered the occurrence of specific medical events using a large Japanese insurance claims database. Their approach predicts whether an event occurs within a given time horizon, whereas our model describes the timing and type of individual claim occurrences through history-dependent conditional intensities, together with the associated claim amounts.

The predictive value of temporal information may also depend on how claim histories are represented and on the prediction task. \citet{faujour2026} compared frequency-based and sequential classifiers using simulated patient trajectories derived from healthcare claims. They found that sequential methods benefited from information about event order, while the inclusion of explicit timing information provided limited additional improvements for their classification task. These findings suggest that the benefits of incorporating detailed temporal information cannot be taken for granted. In our framework, event times enter the specification of the conditional claim intensities, allowing the distribution of future occurrences to depend on the timing and sequence of previous claims. However, our numerical studies do not separately quantify the predictive contribution of exact event timing relative to event order or aggregated claim histories.

Marked point processes provide a natural framework for jointly modeling the timing and characteristics of healthcare events. \citet{mews2023} applied a Markov-modulated marked Poisson process to physician consultations recorded in German medical claims data. In their model, both the occurrence intensity and the mark distribution depend on an unobserved disease state, allowing temporal dependence between consultations to arise through changes in the underlying health status. More flexible representations of event dependence have been developed in the literature on neural temporal point processes. \citet{mei2017} introduced a neural self-modulating point process, while \citet{zuo2020} proposed an attention-based approach to modeling event sequences. These models allow conditional event intensities to depend on representations of preceding events without imposing a fixed parametric structure on their temporal dependence. In a healthcare setting, \citet{enguehard2020} applied neural temporal point processes to electronic health records, jointly modeling the timing and characteristics of medical encounters.

Building on this literature, we develop a neural marked point-process model for individual health insurance claims in which a Transformer representation of past claim histories determines type-specific conditional intensities, while a separate component models conditional claim severities. Unlike the latent-state approach of \citet{mews2023}, our model does not specify an underlying disease process through a finite number of unobserved states, but learns associations between previous claims and subsequent claim occurrences directly from their observed histories. Particular attention is given to features of insurance data, including differences in policy exposure, reporting delays and changes in medical costs over calendar time. These aspects have a long tradition in actuarial modeling, notably in the marked-Poisson framework of \citet{norberg1993} and the reporting-delay analysis of \citet{verrall2016}. Our contribution is to incorporate these elements into a neural marked point-process framework and examine its ability to reproduce and predict claim frequencies, severities and their dependence structure. We evaluate the model using simulated and empirical data and compare its predictive performance with two generalized linear model benchmarks.
\section{The neural marked point process model}\label{sec:model}

We work on a filtered probability space $(\Omega,\mathcal F,(\mathcal F_t)_{0\le t\le \tau},\mathbb P)$ satisfying the usual conditions, where $\tau>0$ denotes a common valuation date. We consider a portfolio of $n$ insured lives, indexed by $i=1,\ldots,n$, and distinguish $K$ claim types, indexed by $k=1,\ldots,K$. Insured $i$ is observed over an exposure interval $[o_i,e_i]\subseteq[0,\tau]$, where $o_i$ and $e_i$ denote the beginning and end of observed coverage, respectively. Coverage may terminate through death, lapse or administrative censoring.

The complete claim history of insured $i$ consists of marked events $(T_{ij},K_{ij},C_{ij},D_{ij})$, indexed by $j$ in order of occurrence. Here, $T_{ij}\in[o_i,e_i]$ denotes the occurrence or treatment time, $K_{ij}\in\{1,\ldots,K\}$ the claim type, $C_{ij}>0$ the claim amount and $D_{ij}>0$ the reporting delay. The corresponding reporting time is $R_{ij}=T_{ij}+D_{ij}$.

For each insured $i$ and claim type $k$, we define the counting process
\begin{equation}\label{eq:counting}
N_{ik}(t)=\sum_j\mathbf 1_{\{T_{ij}\le t,\ K_{ij}=k\}},
\qquad 0\le t\le\tau,
\end{equation}
where the sum extends over all claim occurrences during the exposure interval. Thus, $N_{ik}(t)$ counts the number of type-$k$ claims incurred by insured $i$ up to time $t$.

We interpret $(\mathcal F_t)_{0\le t\le\tau}$ as the full-information filtration, containing the claim occurrences and associated marks up to time $t$, together with the relevant baseline information. We assume that the counting processes are adapted, non-explosive and admit nonnegative, predictable conditional intensities $\lambda_{ik}(t)$ satisfying the required local integrability conditions. In particular, the processes
\begin{equation}\label{eq:martingale}
M_{ik}(t)=N_{ik}(t)-\int_0^t\lambda_{ik}(u)\,du
\end{equation}
are $(\mathcal F_t)$-local martingales. The intensities are understood to vanish outside the corresponding exposure intervals. Under the usual small-interval regularity conditions, the conditional intensity admits the interpretation
\begin{equation}\label{eq:intensity}
\mathbb P\left(
N_{ik}(t+\Delta)-N_{ik}(t)=1
\mid\mathcal F_{t^-}
\right)
=
\lambda_{ik}(t)\Delta+o(\Delta),
\qquad\Delta\downarrow0.
\end{equation}
Thus, $\lambda_{ik}(t)$ describes the instantaneous rate of type-$k$ claim occurrences conditional on the information available immediately before time $t$.

An important distinction arises between the complete occurrence history and the claims observed by the insurer. We write
\begin{equation}\label{eq:hist}
\mathcal H_i(t^-)
=
\bigl\{(T_{ij},K_{ij},C_{ij}):T_{ij}<t\bigr\}
\end{equation}
for the complete occurrence history of insured $i$ immediately before time $t$. At valuation date $\tau$, however, the insurer observes only claims that have been reported by that date. The corresponding set of observations is
\begin{equation}\label{eq:reported}
\mathcal O_i(\tau)
=
\bigl\{(T_{ij},K_{ij},C_{ij},D_{ij}):
T_{ij}\in[o_i,e_i],\ R_{ij}\le\tau\bigr\}.
\end{equation}
We denote the indices of these reported claims by
\[
J_i^\tau
=
\{j:T_{ij}\in[o_i,e_i],\ R_{ij}\le\tau\}.
\]
Consequently, the observed claim history may differ from the complete occurrence history because some claims incurred before valuation have not yet been reported. Moreover, for insured lives whose coverage began before the observation window, earlier claims may be unavailable. The fitted model accounts for the length of this unobserved period through an additional feature but does not reconstruct the missing claims.

In our model, the conditional claim intensities depend on the preceding claim history. To represent histories of varying length and capture dependencies between successive claims, we use a causal Transformer. Its inputs are numerical representations of the recorded claim events, incorporating their timing, type and amount. Specifically, for each recorded claim $(T_{ij},K_{ij},C_{ij})$, we construct the feature vector 
\begin{equation}\label{eq:token}
 v_{ij}=\left(
 \log(T_{ij}-T_{i,j-1}+\varepsilon),\;
 T_{ij}-a_i^{\mathrm{entry}},\;
 \log C_{ij}-M_{K_{ij}}(T_{ij}),\;
 e_{K_{ij}}\right)\in\mathbb R^{K+3}.
\end{equation}
Here $a_i^{\mathrm{entry}}$ is the contract entry date, $T_{i0}
=a_i^{\mathrm{entry}}$ by convention, and $\varepsilon=1/365$ year
ensures that the logarithm is defined for events recorded on the same
date. The vector $e_k\in\mathbb R^K$ has a one in position $k$ and
zeros elsewhere (the \emph{one-hot encoding} of the claim type). The function
$M_k(t)$ denotes the type-specific log medical-cost index, defined below;
subtracting it expresses a historical claim amount relative to the
estimated cost level at its occurrence time. Fixed rescalings of the
continuous coordinates are applied before the following transformation.

Each claim vector is mapped to a $d$-dimensional \emph{embedding} through a learned affine transformation,
\begin{equation}\label{eq:embedding}
z_{ij}=W\widetilde v_{ij}+b,
\qquad
W\in\mathbb R^{d\times(K+3)},\quad b\in\mathbb R^d,
\end{equation}
where $\widetilde v_{ij}$ denotes the rescaled feature vector. 
A further learned embedding $z_0\in\mathbb R^d$ denotes the start of
the recorded history. The same embedding parameters are used for all
insureds.

We use a Transformer encoder to turn the sequence of recorded claims into a
numerical summary of the history available at each point in time
\citep{vaswani2017,zuo2020}. The reason for doing so is that the relevance of
an earlier claim may depend on the claims that followed it. For example, a
hospital admission might be important when interpreting a later outpatient
claim, but less important in a different sequence of events. Instead of
specifying in advance which earlier claims matter, the Transformer learns how
to combine information from the history when the model is trained to predict
claim occurrences.

The central operation is called \emph{self-attention}. Consider the
representation at position $j$, after the $j$th recorded claim. For each
position $\ell\leq j$, the model first calculates a number measuring how
strongly the information at position $\ell$ should contribute to the
representation at $j$. It then uses these numbers to form a weighted average
of information from the current and earlier positions. Thus, attention has
two distinct steps: deciding \emph{where to draw information from}, and
combining the information obtained.

For each claim embedding $z_{ij}\in\mathbb R^d$, the model constructs three vectors, referred to as the \emph{query}, \emph{key} and \emph{value}:
\begin{equation}\label{eq:qkv}
q_{ij}=W_Qz_{ij}+b_Q,\qquad
k_{ij}=W_Kz_{ij}+b_K,\qquad
u_{ij}=W_Vz_{ij}+b_V.
\end{equation}
The query and key vectors, $q_{ij}$ and $k_{ij}$, determine the attention weights assigned to previous claims, while the value vector $u_{ij}$ contains the information combined using these weights.
Their purpose is to determine which positions in the history should be given
more weight. To assess the contribution of an earlier position $\ell$ to the
current position $j$, the model computes the score
\begin{equation}\label{eq:attention_score}
 s_{ij\ell}=\frac{q_{ij}^{\mathsf T}k_{i\ell}}{\sqrt{d_h}}.
\end{equation}
Here $q_{ij}^{\mathsf T}k_{i\ell}$ is the \emph{dot product}: the sum of the
products of corresponding coordinates of the two vectors. A larger score
means that, according to the model's current parameters, position $\ell$ is
more relevant when constructing the representation at $j$. Division by
$\sqrt{d_h}$ controls the magnitude of the scores, where $d_h$ is the number
of coordinates in each query and key vector.

The important point is that the matrices $W_Q$ and $W_K$ are \emph{not}
chosen to measure a predefined similarity between claims. Their entries are
trainable parameters. At the beginning of training, the resulting scores
need not be informative. The query and key projection matrices are learned jointly with the remaining model parameters during training. The attention weights depend on these matrices and influence the conditional claim intensities produced by the neural network. Consequently, the parameters are adjusted to minimize a loss function based on the observed claim histories. In this way, the model learns
which combinations of claim characteristics should give rise to larger or
smaller scores. There are no separately observed ``correct'' attention
weights: the weights are learned indirectly through the prediction task.

For each current position $j$, the scores are converted into nonnegative
weights by exponentiating them and dividing by their sum:
\begin{equation}\label{eq:attention_weights}
 \alpha_{ij\ell}
 =\frac{\exp\!\big(q_{ij}^{\mathsf T}k_{i\ell}/\sqrt{d_h}\big)}
 {\displaystyle\sum_{r=0}^{j}
  \exp\!\big(q_{ij}^{\mathsf T}k_{ir}/\sqrt{d_h}\big)},
 \qquad \ell=0,\ldots,j,
 \qquad \sum_{\ell=0}^{j}\alpha_{ij\ell}=1.
\end{equation}
This normalization is called \emph{softmax}. It makes the weights comparable
within the history available at position $j$: a claim receives a larger share
of the total weight when its score is high relative to the other scores.
Although the weights sum to one, they are not probabilities of future claims
or estimated causal effects.

Once the attention weights have been determined from the query and key vectors, they are used to combine the value vectors. Specifically, the attention output at position $j$ is given by
\begin{equation}\label{eq:attention_output}
a_{ij}=\sum_{\ell=0}^{j}\alpha_{ij\ell}u_{i\ell}.
\end{equation}
Each value vector $u_{i\ell}$ is a learned linear transformation of the embedding at position $\ell$. The attention output $a_{ij}$ is therefore a weighted sum of these transformed representations, where $\alpha_{ij\ell}$ determines the contribution of position $\ell$ to the output at position $j$. Since the value projection parameters $W_V$ and $b_V$ are also learned during training, the model jointly learns the representations being combined and the weights assigned to them.

The restriction $\ell\leq j$ in \eqref{eq:attention_weights} is called
\emph{causal masking}: positions after $j$ are excluded from the calculation
at $j$. It ensures that a history representation cannot depend on claims
recorded later in the sequence. The Transformer performs several such
attention calculations in parallel, called \emph{heads}. Each head has its
own query, key and value parameters and can therefore learn a different way
of weighting and combining the same history. In our implementation, we use four attention heads, each mapping a 64-dimensional input to a 16-dimensional output. The outputs of the four heads are concatenated into a 64-dimensional vector, which is then passed through a learned linear transformation to obtain the final attention output.

The attention calculation is followed by a feed-forward network that further transforms the representation at each position. Unlike attention, which combines information from different positions in the sequence, the feed-forward network processes each position separately. It allows the model to modify the information gathered through attention before passing it to the next layer.

Both the attention calculation and the feed-forward network are combined with two additional operations. First, a \emph{residual connection} adds the input of each transformation to its output, so that the updated representation retains information from before the transformation. Second, \emph{layer normalization} rescales and recenters the coordinates of the resulting representation, helping to stabilize training. Together, these operations form one Transformer encoder layer.

Our model uses two such encoder layers in succession, each with its own learned parameters. The first layer produces updated representations of the claim history, and the second layer further refines these representations using the same sequence of operations. Because causal masking is applied in both layers, the representation at position \(j\) can only incorporate information from positions \(0,\ldots,j\).

We denote the resulting two-layer encoder by \(\operatorname{Enc}^{\mathrm{causal}}_{\phi}\), where \(\phi\) collects its trainable parameters. After applying a final layer normalization, the encoder outputs
\[
(H_{i0},H_{i1},\ldots,H_{im_i})
=
\operatorname{LayerNorm}\!\left[
\operatorname{Enc}^{\mathrm{causal}}_{\phi}
(z_0,z_{i1},\ldots,z_{im_i})
\right].
\]
Each \(H_{ij}\in\mathbb R^{64}\) is a representation of the claim history up to position \(j\). In particular, \(H_{i0}\) is computed from the initial input \(z_0\) alone and represents the state before the first recorded claim.

Two timing details are important for interpreting the conditional intensity. First, although \(H_{ij}\) incorporates information from claim \(j\), the intensity at its occurrence time \(T_{ij}\) is evaluated using \(H_{i,j-1}\). The claim being modeled therefore cannot contribute to its own intensity. Second, the encoder summarizes only the \emph{recorded} claim history. For an insured who entered before the observation window, earlier claims may be unobserved. In that case, setting \(T_{i0}=a_i^{\mathrm{entry}}\) in \eqref{eq:token} is a convention for constructing the first recorded token, and its first coordinate does not necessarily represent the elapsed time since the insured's most recent claim.

The Transformer encoder produces a representation of the recorded claim history, which is subsequently used to determine the conditional claim intensities. To this end, the encoder output is combined with time-dependent characteristics of the insured and the contract.

Recall that $J_i^\tau$ denotes the set of claims of insured $i$ reported by valuation date $\tau$. Let $m_i=|J_i^\tau|$ denote the number of such claims, indexed by $j=1,\ldots,m_i$ in order of occurrence time. We define
\[
j_i(t)=\sum_{r=1}^{m_i}\ind\{T_{ir}<t\}
\]
as the number of recorded claims that occurred strictly before time $t$. The corresponding encoder output $H_{i,j_i(t)}$ depends only on these preceding claims. In particular, if $t=T_{ij}$ and claim occurrence times are distinct, then $j_i(t)=j-1$. Hence, the claim occurring at $t$ does not enter the representation used to calculate its own conditional intensity.

We collect the additional information available at time $t$ in a
feature vector $q_i(t)$. This vector contains the insured's attained
age, contract duration, logarithmic time since the most recent
recorded claim, sine and cosine terms describing the position
within the calendar year, the length of the unobserved pre-sample
period, and the available covariates. The sine and cosine terms
allow the intensity to vary smoothly over the calendar year,
while the time since the last claim provides information about
the elapsed time that is not explicitly contained in the fixed
history representation. The vector $q_i(t)$ should not be confused with
the query vectors $q_{ij}$ introduced in the attention mechanism.

The two sources of information are combined by a multilayer
perceptron, denoted by $g_\theta$. This is a neural network that
successively applies learned linear transformations and nonlinear
functions to its input. Its parameters are estimated jointly with
those of the Transformer. The network produces one numerical
output for each claim type $k$. These outputs can take any real
value and therefore cannot be used directly as intensities,
which must be nonnegative.

For this reason, we apply the softplus transformation
$\softplus(z)=\log(1+e^z)$, which maps any real number to a
positive value. Suppressing for the moment the distinction
between true and reconstructed histories, the resulting
neural intensity for claim type $k$ is
\begin{equation}\label{eq:neuralint}
\begin{aligned}
 \lambda_{ik}^{0}(t;\theta,M)
 &=b_k^{\mathrm{bal}}\exp\{x_i^{\mathsf T}\gamma_k\}\,
 \Bigl[
 \softplus\bigl\{
 g_{\theta,k}(H_{i,j_i(t)},q_i(t))
 \bigr\}+10^{-6}
 \Bigr],\\
 \softplus(z)&=\log(1+e^z).
\end{aligned}
\end{equation}

The neural network $g_{\theta,k}$ determines how the recorded
claim history and the time-dependent characteristics jointly
contribute to the intensity of claim type $k$. Because the
transformations are nonlinear and their parameters are learned
from the data, the model can represent relationships that are
not specified in advance. In particular, the contribution of
the recorded claim history need not be additive or identical
across claim types.

The remaining factors in \eqref{eq:neuralint} serve separate
purposes. The term $\exp\{x_i^{\mathsf T}\gamma_k\}$ represents
a multiplicative effect of the covariates $x_i$, with
claim-type-specific coefficients $\gamma_k$. The constant
$10^{-6}$ provides a small positive numerical offset. Finally, $b_k^{\mathrm{bal}}$ is a positive calibration factor
applied after model estimation. Its purpose is to ensure that
the aggregate predicted number of claims of type $k$ matches
the corresponding observed total in the training sample. This
adjustment is motivated by the \emph{balance property} in
insurance pricing, which requires aggregate fitted claims
experience to agree with aggregate observed claims experience.
As discussed by \citet{wuthrich2022} and
\citet{lindholm2026}, this property is not
automatically satisfied by flexible machine learning models
and may require an additional calibration step. The factor
$b_k^{\mathrm{bal}}$ restores aggregate balance without
changing the relative intensities predicted by the fitted
model. It is therefore a calibration factor rather than a
parameter describing the dependence between successive claims.

The model also allows for unobserved differences in claim
frequency between insureds through type-specific multiplicative
frailties,
\begin{equation}\label{eq:frailty}
 \lambda_{ik}(t\mid Z_{ik})
 =Z_{ik}\lambda_{ik}^{0}(t),\qquad
 Z_{ik}\sim\GammaD(a_k,a_k),\qquad
 a_k=\phi_k^{-1},
\end{equation}
where the Gamma distribution is parameterized by shape and
rate, so that $\E Z_{ik}=1$ and $\Var(Z_{ik})=\phi_k$.
The frailty $Z_{ik}$ is a positive, unobserved quantity
that remains constant for insured $i$ over the observation
period. An insured with $Z_{ik}>1$ has a higher intensity
of claims of type $k$ than would be predicted from the
observed characteristics and claim history alone, whereas
$Z_{ik}<1$ corresponds to a lower intensity.

The frailties are assumed independent across claim types
and insureds. This differs from the data-generating model,
which additionally contains a frailty shared across claim
types and therefore allows unobserved differences in claim
frequency to affect several claim types simultaneously.

The distinction between the frailty and the neural history
representation is important. The Transformer adjusts the
intensity according to the sequence of previously recorded
claims, whereas the frailty accounts for persistent
differences in claim frequency that are not explained by
the recorded information. Both mechanisms can therefore
contribute to higher claim frequencies among insureds
who have already experienced many claims. For example,
repeated claims may indicate a persistently higher
individual risk, but they may also provide information
about changes in risk associated with the preceding
claim history. Consequently, observed correlations
between claim counts do not, by themselves, determine
how much of the dependence should be attributed to
unobserved heterogeneity and how much to the recorded
claim history.

The dependence of the conditional intensity on the recorded
claim history allows the model to represent how previous
claims are associated with subsequent claim occurrences.
In particular, the occurrence of a claim changes the
history representation produced by the Transformer and
may therefore change the output of $g_{\theta,k}$.
This change can be positive or negative and may differ
across claim types. For example, a hospital admission
may be associated with a higher subsequent intensity
of outpatient claims, while another type of event may
be associated with a lower intensity. An increase in
intensity following a claim of the same type is commonly
called self-excitation, whereas an increase following
a claim of a different type is called cross-excitation.
The model can represent both forms of dependence, as
well as decreases in intensity, without imposing their
existence or specifying their magnitude in advance.

In addition to modeling claim occurrences, the model
specifies a conditional distribution for claim amounts.
In point-process terminology, the claim amount is
called a mark: it is an additional quantity observed
when an event occurs. We model the logarithm of the
claim amount after subtracting the type-specific
medical inflation adjustment $M_k(t)$. For claim $j$
of insured $i$, this gives
\[
Y_{ij}
=
\log C_{ij}-M_{K_{ij}}(T_{ij}),
\]
where $C_{ij}>0$ denotes the claim amount and
$K_{ij}$ its type. The logarithmic transformation
allows the model to describe positive claim amounts
on an unrestricted numerical scale, while
$M_k(t)$ accounts for the cumulative effect of
medical inflation on claim amounts of type $k$
up to time $t$. Subtracting this adjustment
removes the specified medical inflation component
from the logarithmic claim amounts.

The remaining variation is modeled using a mixture
of three Gaussian distributions. Rather than assuming
that all transformed claim amounts follow a single
Gaussian distribution, the mixture allows the
conditional density to combine components with
different locations and dispersions. This provides
additional flexibility for describing asymmetric
distributions and observations in the tails.

Conditional on the claim type and the information
available immediately before the claim, the density
of the transformed amount is
\begin{equation}\label{eq:severity}
 p_{\psi}(y\mid H_{i,j-1},q_i(T_{ij}),K_{ij}=k)
 =
 \sum_{r=1}^{3}
 \pi_{ikr}\,
 \varphi(y;\mu_{ikr},\sigma_{ikr}^2),
 \qquad
 \sum_{r=1}^{3}\pi_{ikr}=1,
\end{equation}
where $\varphi(y;\mu,\sigma^2)$ denotes the Gaussian
density with mean $\mu$ and variance $\sigma^2$.
The quantities $\pi_{ikr}$ are nonnegative mixture
weights, while $\mu_{ikr}$ and $\sigma_{ikr}>0$
are the mean and standard deviation of component
$r$. The mixture weights determine the relative
contribution of each component to the conditional
density.

These quantities are produced by a separate
multilayer perceptron with parameters $\psi$.
The network uses the pre-event history representation,
the time-dependent features and the claim type to
determine the mixture distribution. Its outputs
include three numerical values, called logits,
which are transformed into nonnegative weights
summing to one using the softmax transformation.
The network also produces the component means
and positive standard deviations, with
type-specific location intercepts. Consequently,
the conditional distribution of claim amounts
can vary with the recorded claim history and
the insured's characteristics, rather than
remaining fixed within each claim type.

The density in \eqref{eq:severity} describes the
transformed amount $Y$, whereas the observed
quantity is the claim amount $C$. Since
$Y=\log C-M_k(t)$, the corresponding density
of $C$ is obtained by transforming back to
the original monetary scale:
\begin{equation}\label{eq:amountdensity}
 f_C(c\mid t,k,\calH_i(t^-),M)
 =
 \frac{1}{c}\,
 p_{\psi}\!\left(
 \log c-M_k(t)
 \mid H_{i,j_i(t)},q_i(t),k
 \right),
 \qquad c>0.
\end{equation}
The factor $1/c$ arises because the derivative
of $\log c-M_k(t)$ with respect to $c$ is $1/c$.
It ensures that the resulting density integrates
to one on the positive real line. For an observed
claim amount, this factor does not depend on
the neural network parameters, but it is required
for the correct probabilistic specification
of the claim amount distribution.

\section{Reporting delays and calendar-time medical costs}\label{sec:marks}

An insurer observes only claims reported by valuation. Conditional on type $k$ and covariates $x_i$, the fitted reporting delay is assumed to satisfy
\begin{equation}\label{eq:delay}
 \log D_{ij}\mid(K_{ij}=k,x_i)\sim\NN(a_k^D+x_i^{\mathsf T}\beta_D,(s_k^D)^2),
\end{equation}
so for $d>0$ its density and distribution function are
\begin{align}\label{eq:delaycdf}
 f_{D,k}(d\mid x_i)&=\frac{1}{d s_k^D\sqrt{2\pi}}
  \exp\!\left[-\frac{(\log d-a_k^D-x_i^{\mathsf T}\beta_D)^2}{2(s_k^D)^2}\right],\\
 F_{D,k}(d\mid x_i)&=\Phi\!\left(\frac{\log d-a_k^D-x_i^{\mathsf T}\beta_D}{s_k^D}\right),\qquad F_{D,k}(d\mid x_i)=0\ \text{for }d\le0.\notag
\end{align}
Delays are conditionally independent across claims in the fitted specification. A delay is observed only if $D_{ij}\le\tau-T_{ij}$, so the observed delays form a truncated sample. The joint occurrence--delay likelihood below accounts for this truncation through the probability that an occurrence is reported by valuation. Fitting an ordinary lognormal density to observed delays in isolation would not be an equivalent procedure.

Medical cost inflation is modeled separately from changes in age or contract duration. On a monthly grid $t_g=g\Delta$, with $\Delta=1/12$ year, let
\begin{equation}\label{eq:costindex}
 M_k(t_g)=m_g+\delta_{kg},\qquad m_0=0,\quad\delta_{k0}=0.
\end{equation}
The common level $m_g$ and its local drift $v_g$ follow
\begin{equation}\label{eq:trend}
 \begin{split}
 m_{g+1}&=m_g+v_g\Delta+\sigma_m\sqrt{\Delta}\,\epsilon_g,\\
 v_{g+1}&=\bar v+\rho(v_g-\bar v)+\sigma_v\xi_g,
 \end{split}
\end{equation}
while type-specific deviations have the analogous dynamics
\begin{equation}\label{eq:devtrend}
 \begin{split}
 \delta_{k,g+1}&=\delta_{kg}+\eta_{kg}\Delta+\sigma_{\delta}\sqrt{\Delta}\,\epsilon_{kg}^{\delta},\\
 \eta_{k,g+1}&=\rho_{\delta}\eta_{kg}+\sigma_{\eta}\xi_{kg}^{\delta}.
 \end{split}
\end{equation}
The innovations are independent standard normal variables, conditional on the hyperparameters. The autoregressive coefficients lie in $(0,1)$, and the initial drifts receive the corresponding stationary Gaussian distributions. Linear interpolation between the grid dates defines $M_k(t)$ for arbitrary treatment times. Weakly informative priors constrain the drift and innovation scales; in particular the common long-run drift is centered at 3\% per year in the fitted configuration.

Fixing $M_k(0)=0$ is necessary because an unrestricted level shift would be confounded with the type-specific severity intercepts. This normalization does not separately identify age, duration and calendar-time effects. Over a six-year observation window, changes in the estimated index may partly reflect the state-space prior. We therefore compare the fitted paths with the known generating paths, while interpreting the displayed posterior bands conditionally on this regularized model.

\section{Likelihood and numerical calibration}\label{sec:likelihood}

We begin with the complete-data likelihood, conditional on the pre-observation state, medical inflation states and frailties. The full claim history $\calH_i^{\mathrm{full}}$ includes all occurrences during coverage, whether reported by the valuation date $\tau$ or not. In this complete-data formulation, latent quantities are treated as observed. Death and lapse are included as competing causes of coverage termination. For insured $i$, the complete-data log-likelihood is
\begin{align}\label{eq:complete}
 \ell_i^{\mathrm{full}}
 ={}&\sum_{j:T_{ij}\in[o_i,e_i]}
 \Big[
 \log\lambda_{iK_{ij}}(T_{ij}\mid\calH_i(T_{ij}^-))
 +\log f_C(C_{ij}\mid T_{ij},K_{ij},\calH_i(T_{ij}^-),M)
 \notag\\
 &\hspace{38mm}
 +\log f_{D,K_{ij}}(D_{ij}\mid x_i)
 \Big]
 -\int_{o_i}^{e_i}\sum_{k=1}^K
 \lambda_{ik}(u\mid\calH_i(u^-))\dd u
 +\ell_i^{\mathrm{exit}}.
\end{align}
The sum gives the contributions of claim times, types, amounts and reporting delays; the integral is the cumulative claim intensity over the exposure period, accounting also for intervals without occurrences. The exit term uses the cause-specific death and lapse hazards, $\mu_i^d$ and $\mu_i^{\ell}$:
\begin{equation}\label{eq:exit}
 \ell_i^{\mathrm{exit}}
 =\ind\{E_i=\mathrm{death}\}\log\mu_i^d(e_i)
 +\ind\{E_i=\mathrm{lapse}\}\log\mu_i^{\ell}(e_i)
 -\int_{o_i}^{e_i}\{\mu_i^d(u)+\mu_i^{\ell}(u)\}\dd u.
\end{equation}
The hazard term applies to the observed exit cause; the integral accounts for survival without exit up to $e_i$. If coverage is still active at the observation boundary, the exit time is right-censored and only the integral remains. Death and lapse parameters are first estimated separately by maximizing the summed exit log-likelihood \eqref{eq:exit} with the same smoothness penalty used below. This first stage uses full-batch Adam for 400 steps and deterministic quadrature on a 48-point grid. The parameters are then refined jointly within \eqref{eq:elbo}, using the base learning rate without weight decay.

To accommodate reporting delays, first suppose that, conditional on covariates and latent risk factors, claims of type $k$ follow a Poisson process with history-independent intensity $\lambda_{ik}^{P}(t)$ and independently generated reporting delays. A claim occurring at $t$ is reported by $\tau$ with probability $F_{D,k}(\tau-t\mid x_i)$. Independent retention of each occurrence with this probability preserves the Poisson process, a result known as Poisson thinning. Thus, reporting delays can be handled without explicitly enumerating the unreported claims. The reported-claim intensity and its integral are
\begin{equation}\label{eq:thinning}
\begin{aligned}
 \lambda_{ik}^{\mathrm{rep},\tau}(t)
 &=\lambda_{ik}^{P}(t)F_{D,k}(\tau-t\mid x_i),\\
 A_{ik}^{\mathrm{rep},\tau}
 &=\int_{o_i}^{e_i}\lambda_{ik}^{P}(u)
 F_{D,k}(\tau-u\mid x_i)\dd u.
\end{aligned}
\end{equation}
Here $A_{ik}^{\mathrm{rep},\tau}$ is the expected number of type-$k$ claims reported by $\tau$. Writing $J_i^\tau$ for the set of such claims, the likelihood uses the observed event densities and subtracts the expected number of reported events:
\begin{align}\label{eq:poisLL}
 \ell_i^{P,\tau}
 ={}&\sum_{j\in J_i^{\tau}}
 \Big[
 \log\lambda_{iK_{ij}}^{P}(T_{ij})
 +\log f_{D,K_{ij}}(D_{ij}\mid x_i)
 +\log f_C(C_{ij}\mid T_{ij},K_{ij},M)
 \Big]\notag\\
 &-\sum_{k=1}^{K}A_{ik}^{\mathrm{rep},\tau}.
\end{align}
The exit contribution \eqref{eq:exit} is added separately. Thinning is exact conditional on the latent factors; integrating them out need not preserve the Poisson property. This construction is standard in individual-claim reserving \citep{norberg1993,verrall2016,boumezoued2017}.

For the neural model, however, the Transformer intensity depends on previous claims. Unreported occurrences may thus affect later intensities, so independent thinning no longer yields the exact observed-data likelihood. For example, an unreported hospitalization can affect the intensity of subsequent outpatient claims. Exact inference would integrate over the unknown number, times, types, amounts and delays of unreported claims, as well as latent frailties and any missing pre-observation history. Since these quantities also influence later intensities, this integration is generally high-dimensional.

We instead use the claim history reconstructed retrospectively from reports available at $\tau$:
\begin{equation}\label{eq:retrospective}
 \calH_i^{\tau}(t^-)
 =\bigl\{(T_{ij},K_{ij},C_{ij}):R_{ij}\leq\tau,\;T_{ij}<t\bigr\}.
\end{equation}
Claims are ordered by occurrence time. The reconstructed history excludes claims unreported at $\tau$, but can include claims reported after $t$; hence it is neither the full occurrence history nor the information available in real time at $t$. We evaluate both the intensity and severity model against this same history. Define
\[
 \widetilde\lambda_{ik}(t)
 =\lambda_{ik}\bigl(t\mid\calH_i^\tau(t^-)\bigr),\qquad
 \widetilde f_{C,ij}
 =f_C\bigl(C_{ij}\mid T_{ij},K_{ij},\calH_i^\tau(T_{ij}^-),M\bigr).
\]
The delay-adjusted working log-likelihood is then
\begin{align}\label{eq:workLL}
 \widetilde\ell_i
 ={}&\sum_{j\in J_i^\tau}
 \Big[
 \log\widetilde\lambda_{iK_{ij}}(T_{ij})
 +\log\widetilde f_{C,ij}
 +\log f_{D,K_{ij}}(D_{ij}\mid x_i)
 \Big]\notag\\
 &-\int_{o_i}^{e_i}\sum_{k=1}^{K}
 \widetilde\lambda_{ik}(u)F_{D,k}(\tau-u\mid x_i)\dd u
 +\ell_i^{\mathrm{exit}}.
\end{align}
This applies the Poisson delay adjustment to intensities evaluated on reconstructed histories. It agrees with \eqref{eq:poisLL} plus the exit term under the conditional Poisson model with history-independent claim marks. With reporting delays and history-dependent intensities, however, it replaces integration over unreported occurrences with a fixed retrospective history, so its maximizer is a working-likelihood estimator rather than an exact observed-data maximum-likelihood estimator.

Type-specific gamma frailties \eqref{eq:frailty} account for residual heterogeneity. Their multiplicative form, $\lambda_{ik}(t)=Z_{ik}\lambda_{ik}^{0}(t)$, permits analytical integration within the working likelihood. Let $n_{ik}$ be the number of reported type-$k$ claims and define the delay-adjusted baseline cumulative intensity
\begin{equation}\label{eq:Ak}
 A_{ik}(\theta,M)
 =\int_{o_i}^{e_i}
 \lambda_{ik}^{0}\bigl(u\mid\calH_i^\tau(u^-)\bigr)
 F_{D,k}(\tau-u\mid x_i)\dd u.
\end{equation}
The frailty-dependent occurrence contribution is proportional to $Z_{ik}^{n_{ik}}\exp(-Z_{ik}A_{ik})$. Integrating over the gamma distribution yields
\begin{align}\label{eq:gammamarg}
 \widetilde\ell_i^{\mathrm{occ}}
 ={}&\sum_{j\in J_i^\tau}
 \log\widetilde\lambda_{iK_{ij}}^{0}(T_{ij})
 \notag\\
 &+\sum_{k=1}^{K}
 \Big[
 \log\Gamma(n_{ik}+a_k)
 -\log\Gamma(a_k)
 +n_{ik}\log\phi_k
 \notag\\
 &\hspace{33mm}
 -(n_{ik}+a_k)\log(1+\phi_k A_{ik})
 \Big].
\end{align}
Here $\widetilde\lambda^0$ is the baseline neural intensity evaluated on reconstructed histories. This marginalization avoids estimating a separate frailty for every insured and claim type. The marginalization is analytical conditional on the reconstructed history, but does not remove any approximation due to unreported claims. Reporting-delay, severity and exit terms complete the calibration criterion.

Because the Transformer intensity varies with time and claim history, $A_{ik}$ is approximated by Monte Carlo quadrature. For each insured in a minibatch, draw $Q=32$ independent $U_{ir}\sim\mathrm{Unif}[o_i,e_i]$ and set
\begin{equation}\label{eq:mc}
 \widehat A_{ik}
 =\frac{e_i-o_i}{Q}
 \sum_{r=1}^{Q}
 \lambda_{ik}^{0}\bigl(U_{ir}\mid\calH_i^\tau(U_{ir}^-)\bigr)
 F_{D,k}(\tau-U_{ir}\mid x_i).
\end{equation}
For fixed parameters and history, $\widehat A_{ik}$ is unbiased for $A_{ik}$, although substituting it into the nonlinear expression \eqref{eq:gammamarg} introduces log-likelihood bias. To reduce computation, a single causal Transformer pass per insured and minibatch step produces the embeddings for all observed history prefixes. A sorted search assigns each quadrature point to its prefix, so only the smaller intensity network must be evaluated at the integration times. Embeddings are not cached across optimization steps, since the Transformer weights change.

Medical inflation is represented by latent monthly levels and drifts in $S$, excluding states fixed for identification. We approximate their joint distribution by a factorized Gaussian,
\begin{equation}\label{eq:var}
 q_{\omega}(S)
 =\prod_{r=1}^{\dim S}
 \NN\bigl(S_r;m_r^{(q)},(s_r^{(q)})^2\bigr),
\end{equation}
where $\omega$ denotes the variational means and standard deviations. The factorization simplifies inference but cannot represent posterior dependence between inflation states. Let $\vartheta$ comprise the neural, reporting-delay, frailty-scale and inflation state-space parameters. We jointly maximize
\begin{equation}\label{eq:elbo}
 \mathcal J(\vartheta,\omega)
 =\E_{q_\omega}\left[
 \sum_{i=1}^{n}\widetilde\ell_i(\vartheta,S)
 +\log p_{\vartheta}(S)\right]
 +\mathcal H(q_\omega)
 +\log p(\vartheta_{\mathrm{state}})
 -\kappa\mathcal P_{\mathrm{exit}}.
\end{equation}
The expectation combines the working likelihood with the inflation state-space density; $\mathcal H(q_\omega)$ is variational entropy. The remaining terms impose a prior on state-space parameters and penalize excessive variation in death and lapse baseline hazards, with penalty strength $\kappa$.

We estimate the expectation with one inflation-state draw per minibatch and scale the minibatch likelihood to the portfolio. Historical claim tokens use the posterior-mean inflation index with gradients stopped, whereas the severity likelihood uses sampled states. Thus, estimation combines variational inference for inflation with optimization of the neural working likelihood.

The reporting-delay specification differs between the numerical studies. In the MEPS application, reporting delays are disabled ($\texttt{delay=False}$): $F_{D,k}\equiv1$, the $\log f_{D,k}$ terms vanish, and the reconstructed history coincides with the full observed claim history within the observation window. The approximation involving unreported claims is therefore unnecessary for MEPS, although pre-observation history may remain unobserved. In the other numerical study, reporting delays are enabled ($\texttt{delay=True}$), so the delay terms and thinning adjustment remain active. For that study, the reconstructed history can omit claims not yet reported by $\tau$, and the history-dependent working-likelihood approximation described above is required. Numerical quadrature and variational inference introduce further approximations in the calibration procedure.
 
\section{Simulation design}\label{sec:sim}

The simulation consists of $n=10{,}000$ synthetic health insurance
contracts observed over six years, beginning in January
2021, with random seed 1. Of these contracts, 45\% are
already in force at the beginning of the observation
period, while the remaining 55\% enter during the
six-year period. Approximately 15\% of entrants are
children, and the remaining entry ages range from
18 to 80 years. The simulated portfolio contains
three tariff classes, with tariff selection depending
on age. The tariff class is used in generating the
data but is not supplied to the fitted neural model.
We distinguish four types of claims: inpatient,
outpatient, dental and pharmacy.

The data are generated from a model specified
independently of the neural network. This allows us
to evaluate how well the fitted model reproduces
features of claim histories without being given
the mechanism that produced them. The generating
model incorporates age- and duration-dependent
claim frequencies, unobserved differences in
individual risk, and dependence between successive
claims. Claim amounts, reporting delays, medical
inflation, mortality and lapse are also simulated.

Claims are generated using type-specific intensities
that combine a baseline frequency with an additional
contribution from previous claims. The baseline
depends on the insured's age, policy duration and
tariff class, as well as calendar time and
unobserved individual risk factors. In particular,
insured $i$ has a common multiplicative frailty
$Z_{i0}$ and a separate frailty $Z_{ik}$ for each
claim type $k$. All frailties have gamma
distributions with mean one and are mutually
independent. The common frailty has variance
$0.3$ and affects the baseline frequencies of
all four claim types. The type-specific frailties
have variances $0.8$, $0.25$, $0.5$ and $0.25$
for inpatient, outpatient, dental and pharmacy
claims, respectively. They allow additional
persistent differences between insureds within
each claim type.

Conditional on these frailties, the generating
intensity for claims of type $k$ is
\begin{equation}\label{eq:true}
\begin{aligned}
 \lambda_{ik}^{*}(t)
 ={}&
 \lambda_k^{0}
 A_k\{a_i(t)\}
 S_k\{d_i(t)\}
 R(\mathrm{tariff}_i)
 e^{0.01t}
 Z_{i0}Z_{ik}\\
 &+
 \sum_{j=1}^{K}
 \alpha_{jk}\beta_j
 \sum_{\substack{s\in\mathcal T_{ij}\\s<t}}
 e^{-\beta_j(t-s)},
\end{aligned}
\end{equation}
where $a_i(t)$ denotes attained age,
$d_i(t)$ is policy duration, $t$ is calendar
time measured in years from the simulation
origin, and $\mathcal T_{ij}$ contains the
previous occurrence times of claims of type
$j$ for insured $i$.

The first term in \eqref{eq:true} describes
the baseline claim frequency. The reference
rates $\lambda_k^0$ are $0.10$, $2.0$,
$0.8$ and $2.0$ claims per year for inpatient,
outpatient, dental and pharmacy claims,
respectively. These rates refer approximately
to an insured aged 40 in the middle tariff
class, after the initial selection period.
Age dependence is introduced through
\[
 A_k(a)
 =
 \left\{
 u_k e^{-a/4}
 +e^{s_k(\min(a,95)-40)}
 \right\}
 e^{-c_k\max(a-65,0)}.
\]
The first component allows for elevated
frequencies in early childhood, while the
second describes an increase with age.
The final factor allows frequencies to
decline after age 65, particularly for
dental claims. The parameters governing
these effects differ across claim types.

The duration-dependent factor is
\[
 S_k(d)=1+\sigma_k e^{-d/\ell_k}.
\]
It represents an initial selection effect
that gradually disappears as policy duration
increases. At entry, inpatient and outpatient
frequencies are reduced by 35\% and 15\%,
respectively, while dental frequency is
increased by 50\%, representing an initial
catch-up effect. These differences fade
over approximately one year. The tariff
factor $R$ takes the values $0.85$, $1$
and $1.15$, and the factor $e^{0.01t}$
introduces an annual increase of
approximately 1\% in baseline claim
frequencies. The frailties multiply
this baseline component, but do not
directly multiply the contribution
from previous claims.

The factor $R(\mathrm{tariff}_i)$ accounts for
differences between three artificial tariff
classes assigned at policy entry, with
assignment probabilities depending on entry
age. The classes multiply baseline claim
frequencies by $0.85$, $1$ and $1.15$,
respectively. They also affect claim amounts,
with factors of $0.90$, $1$ and $1.16$,
and lapse hazards, with factors of $1.2$,
$1$ and $0.8$. These classes do not
represent specific insurance products
but introduce systematic differences in
claim experience and policyholder behavior.
Although tariff membership is recorded
in the simulated data, it is deliberately
excluded from the fitted model, so that
the associated differences must be
accounted for indirectly through the
available covariates and claim histories.

The second term in \eqref{eq:true}
introduces dependence between successive
claims. Each previous claim of type $j$
adds a positive contribution to the
subsequent intensity of type $k$, which
decreases exponentially with the time
elapsed since that claim. The coefficient
$\alpha_{jk}$ determines the total
contribution of one type-$j$ event to
the type-$k$ intensity when integrated
over subsequent time, while $\beta_j$
determines how quickly that contribution
decays. The decay rate is $12$ per year
for most claim types, corresponding to
a mean decay time of approximately one
month, and $8$ per year for dental claims.

For example, an inpatient claim contributes
an integrated intensity increment of $1.0$
for outpatient claims, $1.0$ for pharmacy
claims and $0.15$ for subsequent inpatient
claims. An outpatient claim contributes
$0.15$ to the integrated outpatient
intensity and $0.3$ to the pharmacy
intensity. Dental and pharmacy claims
contribute $0.4$ and $0.1$, respectively,
to their own type. These increments
describe the direct contribution of an
individual event, before accounting for
further claims generated through the
same mechanism. The generating intensity
uses all previous claim occurrences,
including claims before the observation
period and claims that have occurred
but have not yet been reported. The
fitted neural model does not observe
these claims when constructing its
history representations. Moreover,
the generating coefficients and
frailties are not supplied during
estimation.

Claims are simulated in weekly steps.
Within each week, the number of claims
of each type is drawn from a Poisson
distribution using the intensity evaluated
at the midpoint of the week. Claims
occurring during a week begin contributing
to subsequent intensities from the
following week onward.

Claim amounts are generated from a severity
model that combines an age- and tariff-dependent
lognormal distribution with medical inflation
and occasional large-claim adjustments.
At age 40, the reference median claim
amounts are approximately $3{,}500$, $90$,
$180$ and $40$ for inpatient, outpatient,
dental and pharmacy claims, respectively,
in the monetary units of the simulation.
For inpatient and pharmacy claims,
occasional multiplicative Pareto factors
generate particularly large amounts.
A small dependence on the common frailty
also introduces an association between
claim frequency and severity: insureds
with higher underlying claim frequencies
may tend to have different claim amounts.
Medical inflation is represented by a
stochastic cost index with a trend centered
near 3\% annually and type-specific
deviations. Consequently, the distribution
of nominal claim amounts can change over
calendar time even when the remaining
severity characteristics are unchanged.

Each claim is also assigned a reporting
delay, generated from a type-specific
lognormal distribution. The median delays
are approximately 40, 20, 30 and 7 days
for inpatient, outpatient, dental and
pharmacy claims, respectively. These
delays determine whether a claim that
has already occurred is visible in the
data at the valuation date. Insurance
coverage may end through death or lapse.
Mortality follows a Gompertz--Makeham-type
hazard, while the lapse hazard depends
on attained age and policy duration.
Together with contract entry and the
administrative observation boundary,
these exit events determine the exposure
intervals over which claims can occur.

Although the simulator records the complete
generated histories, the fitted model uses
only claims that occur during the insured's
exposure interval and have been reported
by the valuation date $\tau$. Claims
incurred during exposure but reported
after $\tau$ are therefore absent from
the calibration data. For contracts
already in force at the beginning of
the observation period, earlier claims
are also unavailable to the fitted model.
These restrictions reproduce two
important features of insurance data:
recent claims may remain unreported,
and the recorded history of an insured
may begin after the actual start of
coverage. Both affect the information
available for reconstructing dependence
between successive claims.

\begin{figure}[!tbp]
 \centering
 \includegraphics[width=0.96\linewidth]{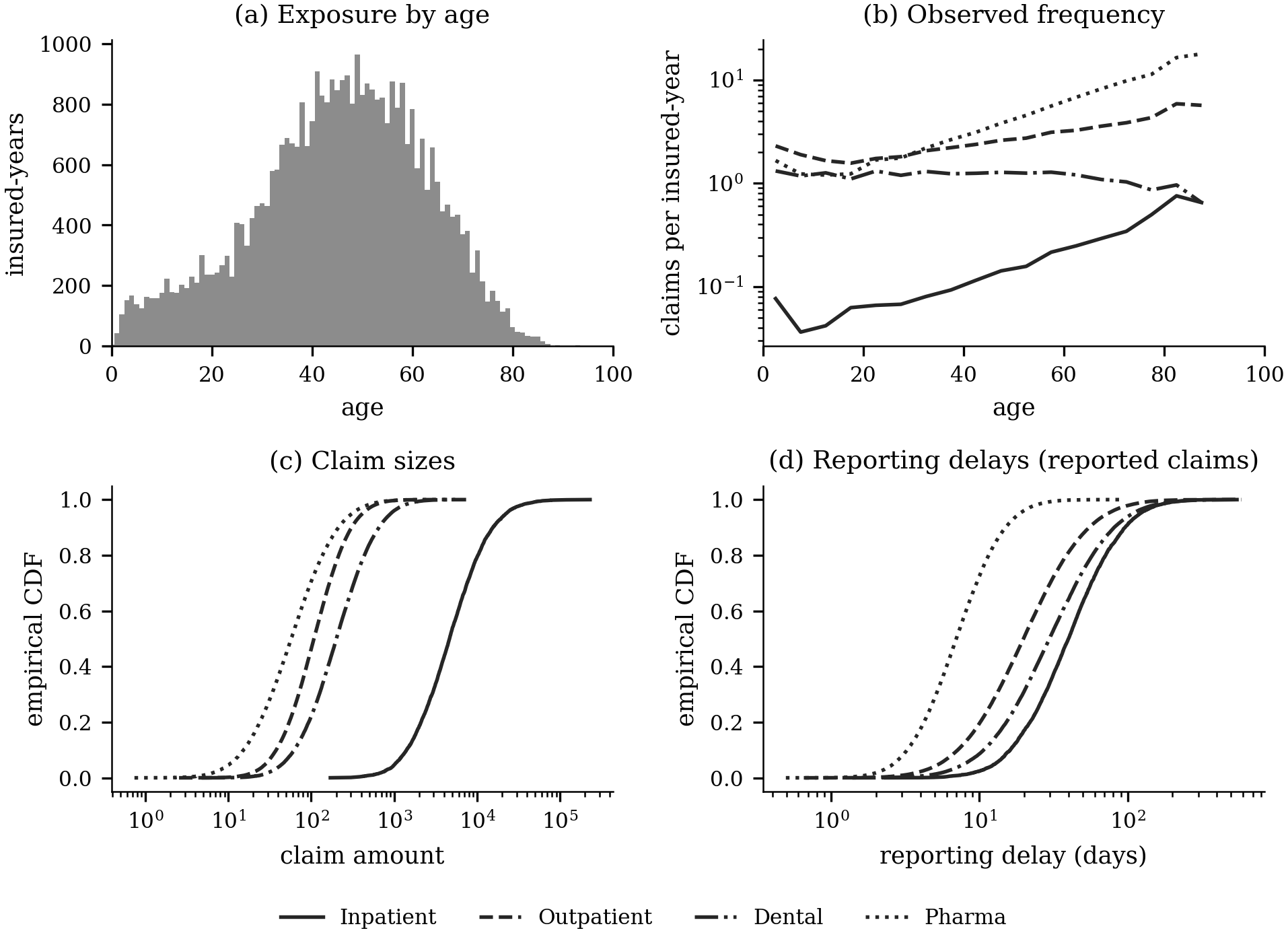}
 \caption{Reported calibration data:
 (a) exposure by age,
 (b) observed claim frequency per insured-year,
 (c) empirical amount distributions, and
 (d) reporting delays.
 Line styles identify the four claim types.}
 \label{fig:portfolio}
\end{figure}

Table~\ref{tab:portfolio} and
Figure~\ref{fig:portfolio} summarize
the resulting reported sample. It
contains $316{,}620$ claims over
$37{,}645$ insured-years. A further
$3{,}791$ claims were incurred during
the exposure periods but had not been
reported by valuation. The observed
claim frequencies exceed the reference
rates $\lambda_k^0$ because the
portfolio includes insureds with
different ages and risk characteristics,
an upward calendar trend, and additional
claims generated through the
history-dependent component.

\begin{table}[!tbp]
\centering\small
\caption{Reported claims in the six-year synthetic
portfolio. Frequencies are per insured-year;
monetary quantities are in the simulation's units.}
\label{tab:portfolio}
\setlength{\tabcolsep}{4.5pt}
\begin{tabular}{lrrrrrrr}
\toprule
Claim type & Claims & Frequency & Mean & Median &
99\% quantile & CV & Delay (days)\\
\midrule
Inpatient & 6,118 & 0.163 & 7,424 & 4,653 &
47,284 & 1.33 & 40\\
Outpatient & 99,728 & 2.649 & 151 & 108 &
718 & 0.98 & 20\\
Dental & 45,974 & 1.221 & 307 & 202 &
1,689 & 1.13 & 30\\
Pharmacy & 164,800 & 4.378 & 97 & 57 &
627 & 1.37 & 7\\
\bottomrule
\end{tabular}
\par\smallskip
\parbox{0.97\linewidth}{\footnotesize
The sample comprises 10,000 contracts and
37,645 insured-years. An additional 3,791
incurred claims are unreported at valuation.
``Delay'' denotes the median observed delay.}
\end{table}

The simulation allows us to examine how the proposed
model performs when claim histories exhibit both
short-term dependence and persistent unobserved
heterogeneity, while reporting delays and incomplete
pre-observation histories restrict the information
available for calibration. Since the experiment uses
one simulated portfolio and one random seed, the
reported results describe performance for this
particular realization rather than variation across
independently generated portfolios.

The parameters of the data-generating model are
illustrative and were not estimated or calibrated
using an observed insurance portfolio. In particular,
they were not obtained from an insurance company,
healthcare provider or any other external data
provider. They were chosen to produce claim frequencies,
amounts and reporting patterns of plausible orders
of magnitude for a European health insurance portfolio,
while ensuring that the features of interest are
sufficiently pronounced to be examined in the
simulation. These include persistent differences
in individual risk, dependence between successive
claims, selection effects after policy entry,
calendar-time trends, dependence between claim
frequency and severity, and reporting delays.
The experiment therefore provides a controlled
setting for assessing whether the proposed model
can recover and predict features of a known
data-generating process. It does not establish
that the generating mechanism, or the numerical
values of its parameters, accurately describe
any particular real-world health insurance portfolio.

\section{Calibration results}\label{sec:results}

For the main comparison, new entrants at the end of the observation window are simulated for their first four policy years under the fitted model and under the generating model. The two sets of simulations use the same entrant population and compare claims before contract terms, at the medical cost level prevailing at the data end. The fitted curves are based on 30 simulation paths with 12 lives per entry age. The uncertainty bands shown in the plots describe simulation variation. In this section, ``true'' refers to the separate generating model, not to observed claims from a real insurer.

\begin{figure}[!tbp]
 \centering\includegraphics[width=0.94\linewidth]{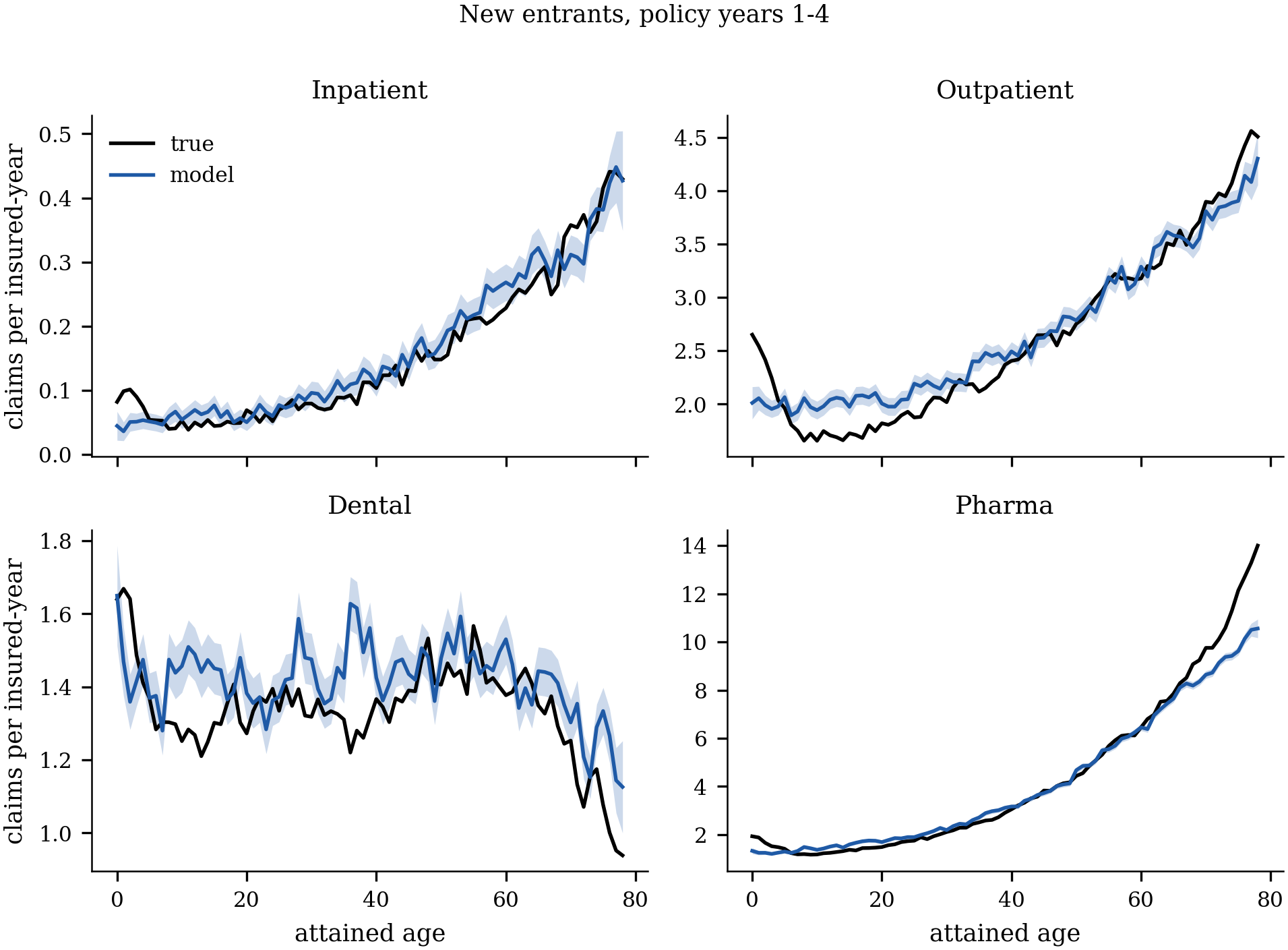}
 \caption{Claim frequencies of new entrants, by attained age, over policy years 1--4. Black: generating model; blue: fitted neural model. Shading reflects Monte Carlo variation in the fitted simulation.}\label{fig:frequency}
\end{figure}

Figure~\ref{fig:frequency} compares claim frequencies
by attained age. The neural model reproduces the
main age-related patterns across the four claim types,
including the substantial increase in inpatient
and pharmacy claim frequencies at older ages.
Differences are more apparent at the extremes of
the age distribution, particularly for pharmacy
claims, where the fitted frequencies underestimate
those of the generating model at the highest ages.
Overall, the comparison suggests that the fitted
model captures much of the variation in claim
frequencies across age groups.

\begin{figure}[!tbp]
 \centering\includegraphics[width=0.98\linewidth]{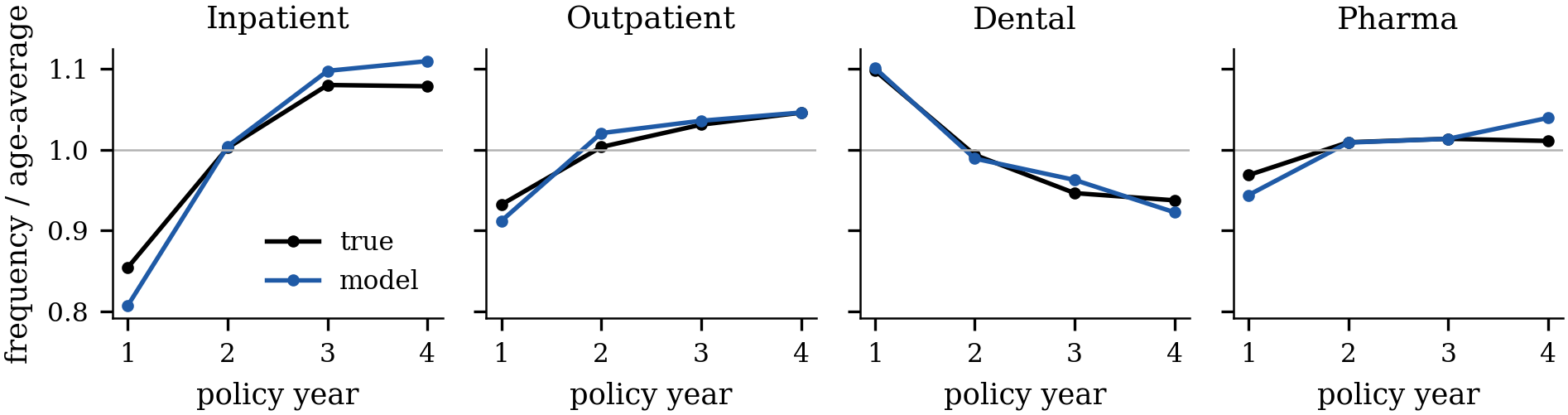}
 \caption{Age-standardized frequency by policy year relative to the age-standardized average. Black: generating model; blue: fitted model.}\label{fig:selection}
\end{figure}

Figure~\ref{fig:selection} compares the duration-dependent claim frequencies under the generating and fitted models. For each policy year, claim frequencies are expressed relative to the corresponding age-standardized mean. The fitted curves broadly recover the initial selection pattern: lower inpatient and outpatient frequency immediately after entry and a temporary elevation of dental claims. Discrepancies remain in the first year and near the end of the four-year period. Recovering these profiles matters because the neural intensity uses both contract duration and event history; a model that captures clustering but misses the duration effect can still give incorrect predictions.

\begin{figure}[!tbp]
 \centering\includegraphics[width=0.96\linewidth]{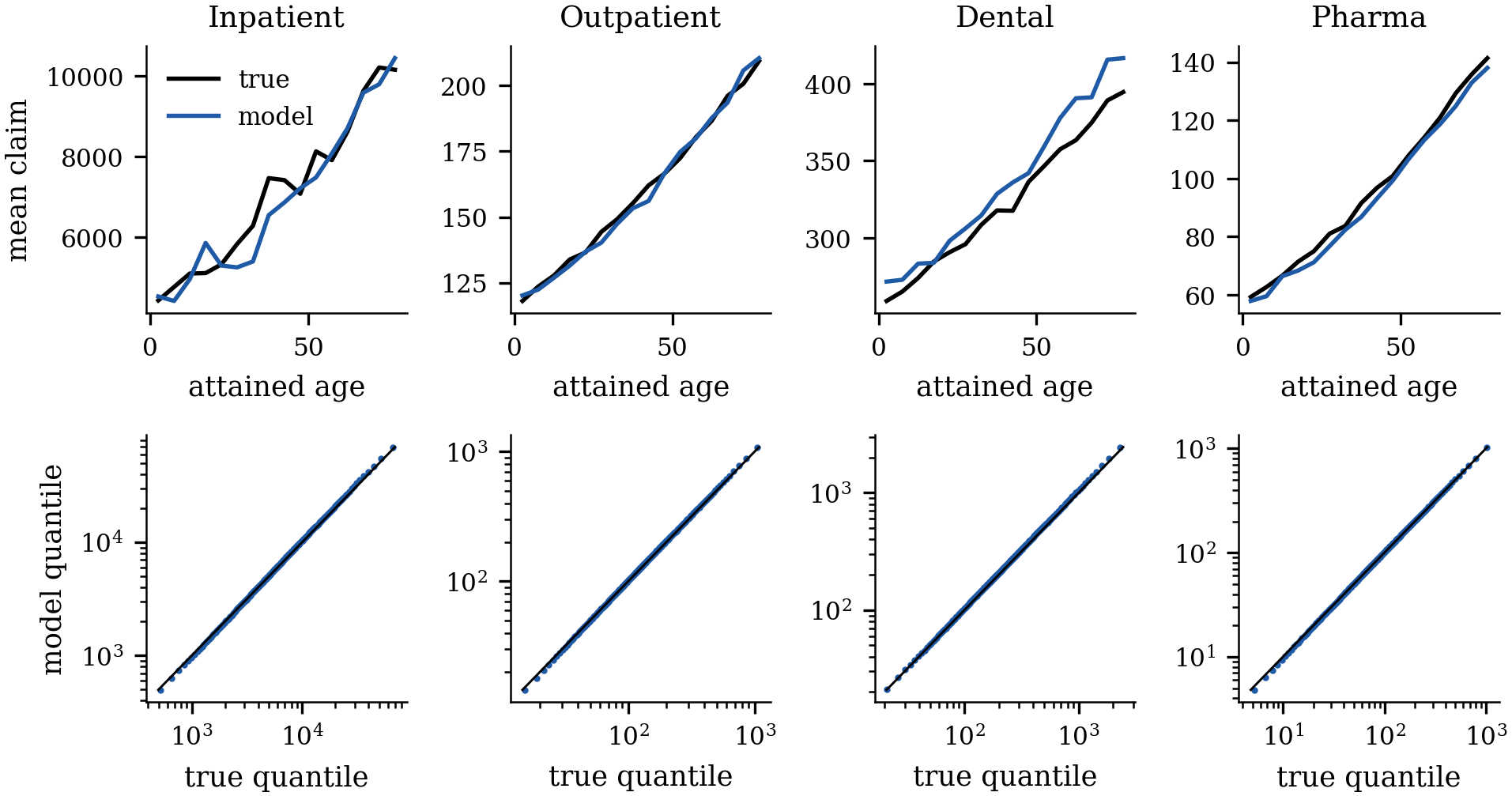}
 \caption{Severity calibration. Upper panels: mean claim amounts by age under the generating (black) and fitted (blue) models. Lower panels: fitted versus generating quantiles on logarithmic axes; the diagonal is exact agreement.}\label{fig:severity}
\end{figure}

The severity comparison in Figure~\ref{fig:severity} considers both conditional means by attained age and marginal quantile--quantile plots. Mean claim amounts generally increase with age, and the fitted curves follow their principal trends. The quantile comparisons are also close across much of the displayed range, including much of the upper tail on the logarithmic scales. This is evidence that the conditional lognormal mixture is flexible enough to approximate the generated amount distributions, which include occasional heavy-tail multipliers. There are still visible deviations in age-specific mean severities, particularly for dental and inpatient claims, and the QQ plots should not be read as a guarantee of accurate extrapolation beyond observed tail quantiles.

\begin{figure}[!tbp]
 \centering\includegraphics[width=0.97\linewidth]{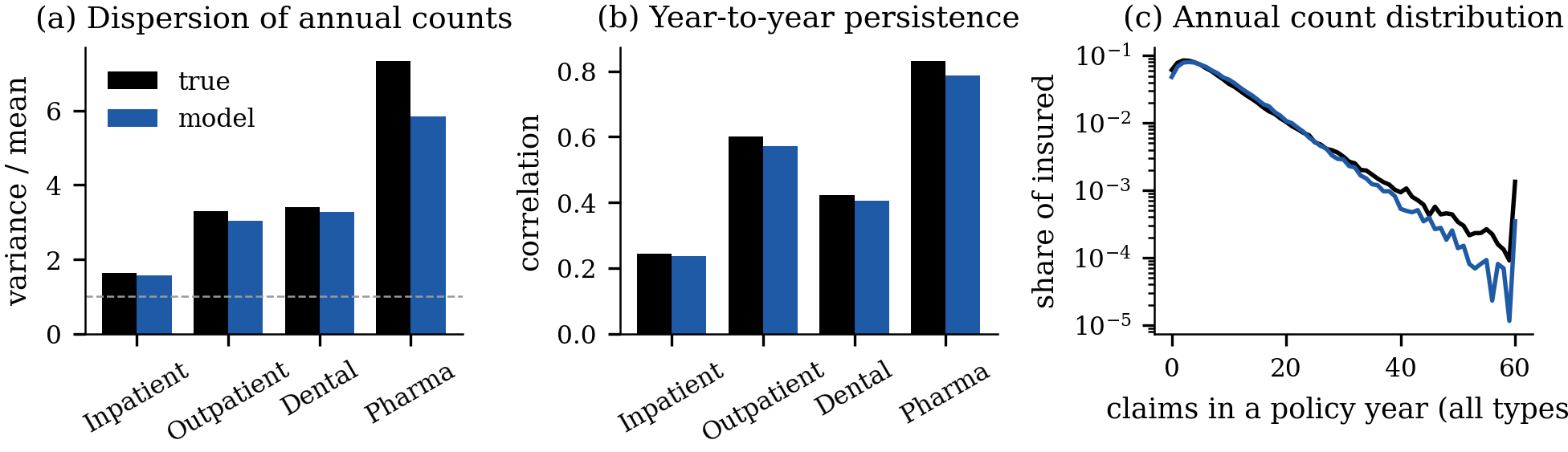}
 \caption{Dependence of annual claim counts in new entrants. (a) Variance-to-mean ratios (one is the Poisson benchmark), (b) year-to-year count correlations, and (c) the annual total-count distribution. Black: generating model; blue: fitted model.}\label{fig:dependence}
\end{figure}

History dependence is not assessed by the marginal mean alone. Figure~\ref{fig:dependence} compares the variance-to-mean ratio of annual claim counts, the within-insured correlation of counts in successive years and the tail of the annual count distribution. The fitted model captures overdispersion and positive persistence in all four categories to a considerable extent. Yet the pharmacy variance-to-mean ratio is substantially below the generating value, and the fitted upper tail of total annual counts is too light. This is important evidence against an unqualified claim of recovery of the joint claim process. Both the generator and the fitted model contain frailty as well as history dependence; these summaries cannot determine whether the neural intensity learned the true excitation kernels.

\begin{figure}[!tbp]
\centering\includegraphics[width=0.97\linewidth]{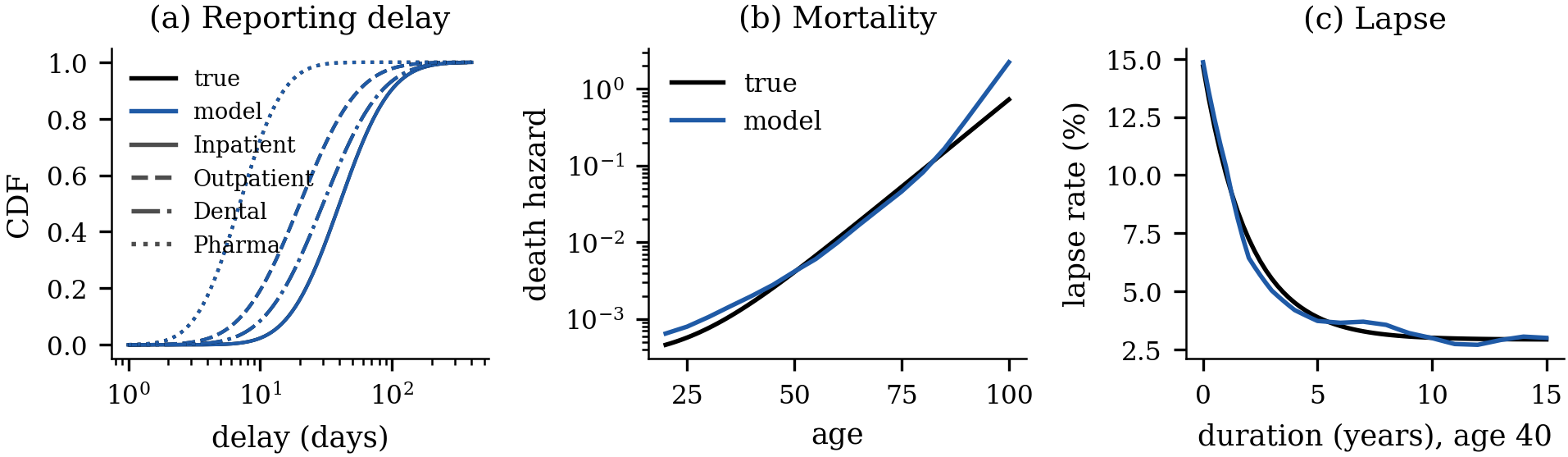}
 \caption{Calibration of the reporting-delay and
competing-risk models in the simulation study.
(a) True and fitted reporting-delay distributions
by claim type.
(b) True and fitted mortality hazards as functions
of attained age.
(c) True and fitted lapse rates by policy duration
for an insured aged 40.}
\label{fig:classical}
\end{figure}

Figure~\ref{fig:classical} compares the fitted reporting-delay, mortality and lapse models with their counterparts in the data-generating process. The estimated reporting-delay distributions closely reproduce the true distributions across all four claim types. Similarly, the fitted mortality hazard captures the age-dependent increase in mortality, although some deviations become apparent at advanced ages. The lapse model also reproduces the pronounced decline in lapse rates during the first years of the contract, with only minor differences at longer durations.

\begin{figure}[!tbp]
 \centering\includegraphics[width=0.97\linewidth]{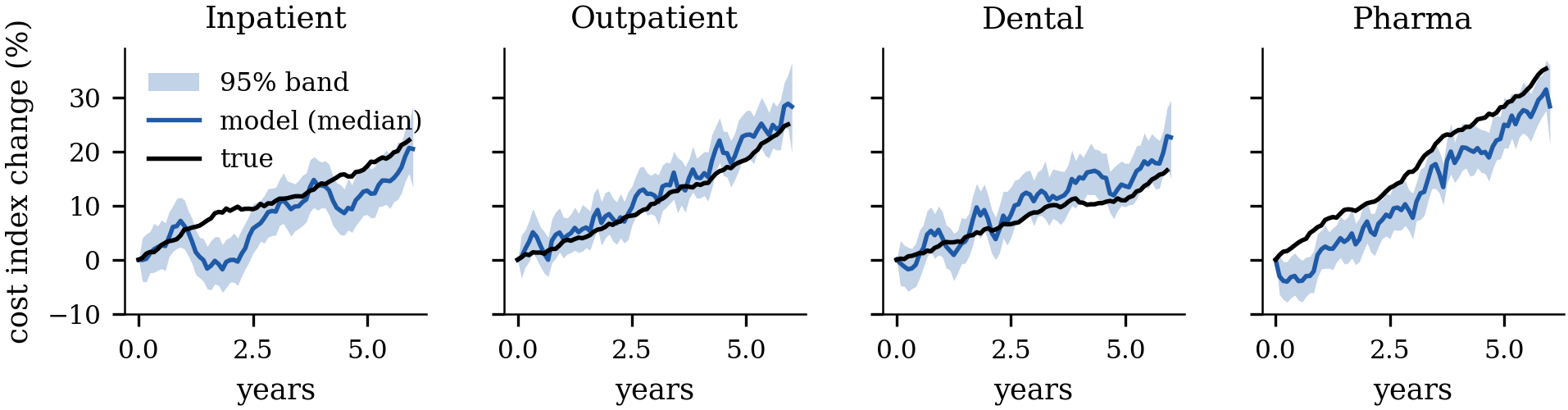}
 \caption{Medical cost indices relative to their initial level. Black: generating path; blue: fitted posterior median; shading: nominal 95\% variational posterior band.}\label{fig:inflation}
\end{figure}

Figure~\ref{fig:inflation} compares the fitted medical
cost indices with the trajectories used to generate
the data. The model captures the general upward trend
over the six-year period, although the agreement
varies across claim types. The outpatient index
follows the generating trajectory relatively closely,
while the fitted pharmacy index tends to underestimate
the increase in medical costs. For inpatient claims,
the fitted index exhibits more pronounced fluctuations,
whereas the dental index generally increases more
rapidly than the generating trajectory. The shaded
bands represent 95\% intervals obtained from the
fitted variational distribution of the inflation
states. These intervals describe uncertainty within
the assumed state-space model and should not be
interpreted as having established 95\% frequentist
coverage. Given the relatively short observation
period and the potential overlap between calendar-time
effects, attained age and policy duration, the
results suggest that the model captures the broad
development of medical costs more reliably than
their precise type-specific trajectories.

Overall, the simulation results indicate that the
proposed model captures several important features
of the underlying data-generating process. The model
reproduces the main patterns in claim frequencies
and severities, while the agreement is less precise
for some claim types and for the development of
medical costs over time. The results also illustrate
how a common modeling framework can accommodate
differences in claim frequency, claim amounts,
reporting delays and dependence between successive
claims. Having examined the model under controlled
conditions, we now turn to an application using
real insurance data.

\section{Application to MEPS-based medical-event data}\label{sec:meps}

The simulation study allows us to compare the fitted model with a known data-generating process. We now examine its performance on medical-event histories from the \emph{Medical Expenditure Panel Survey} (MEPS). We use Panel 24, covering 2019--2022, and distinguish inpatient stays, emergency visits, ambulatory care, dental visits and prescribed medicines. The processed sample contains 5,447 individuals, with 21,001.5 person-years and 436,542 recorded events. Expenditure includes payments from all recorded payers. Because MEPS is a household survey rather than a portfolio of adjudicated insurance claims, it does not provide verified insurer receipt dates. We consequently omit the reporting-delay component in this application, setting $F_{D,k}\equiv 1$, and fix the observation boundary at January 1, 2023. Individuals leaving the survey population are censored; such departures are not interpreted as insurance lapse. We do not use survey weights, so the results refer to the analysis sample rather than population-weighted estimates for the United States. Prescription records do not contain the dates of individual dispensings. Their dates are assigned uniformly within the recorded calendar year, which preserves their annual utilization totals but does not provide evidence about their ordering within the year.

\begin{figure}[!tbp]
\centering
\includegraphics[width=0.46\linewidth]{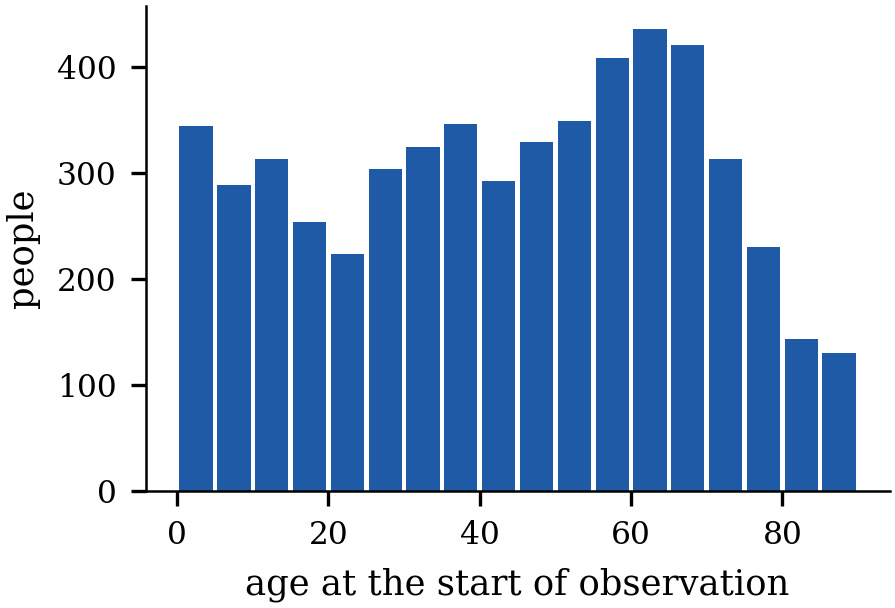}\hfill
\includegraphics[width=0.52\linewidth]{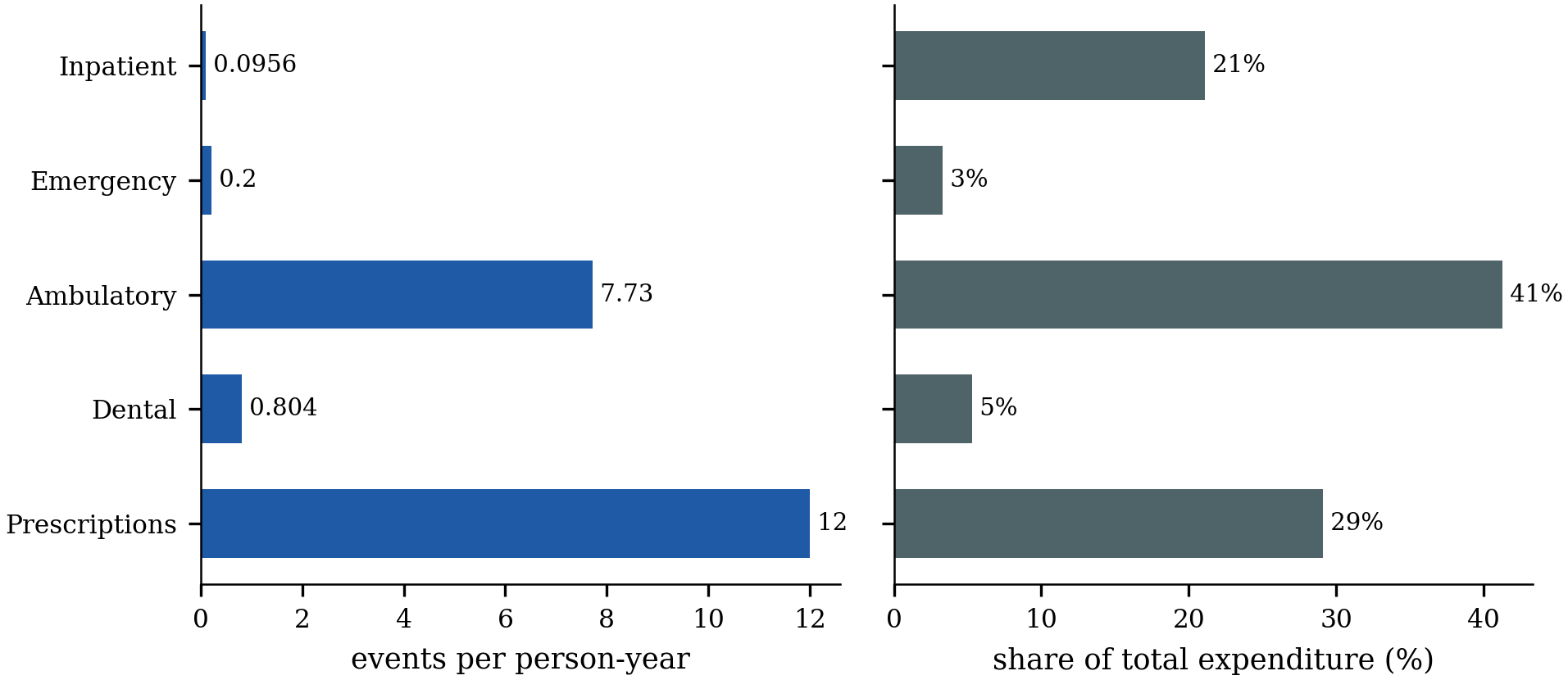}
\caption{Composition of the MEPS analysis sample. Left: age at the start of observation. Right: event frequencies per person-year and shares of total expenditure by type.}\label{fig:meps_population}
\end{figure}

The population and composition of medical expenditure are shown in Figure~\ref{fig:meps_population}. The sample covers children, working-age adults and older individuals. Prescriptions are the most frequent events, at approximately 12 per person-year, followed by ambulatory care at 7.73. Inpatient stays and emergency visits are much less frequent, at 0.0956 and 0.20 per person-year, respectively. The distribution of expenditure is quite different: ambulatory care accounts for 41\% of the total and prescriptions for 29\%, while inpatient stays account for 21\% despite their low frequency. These differences make it important to evaluate both event counts and amounts, rather than relying on aggregate expenditure alone.

\begin{figure}[!tbp]
\centering\includegraphics[width=0.98\linewidth]{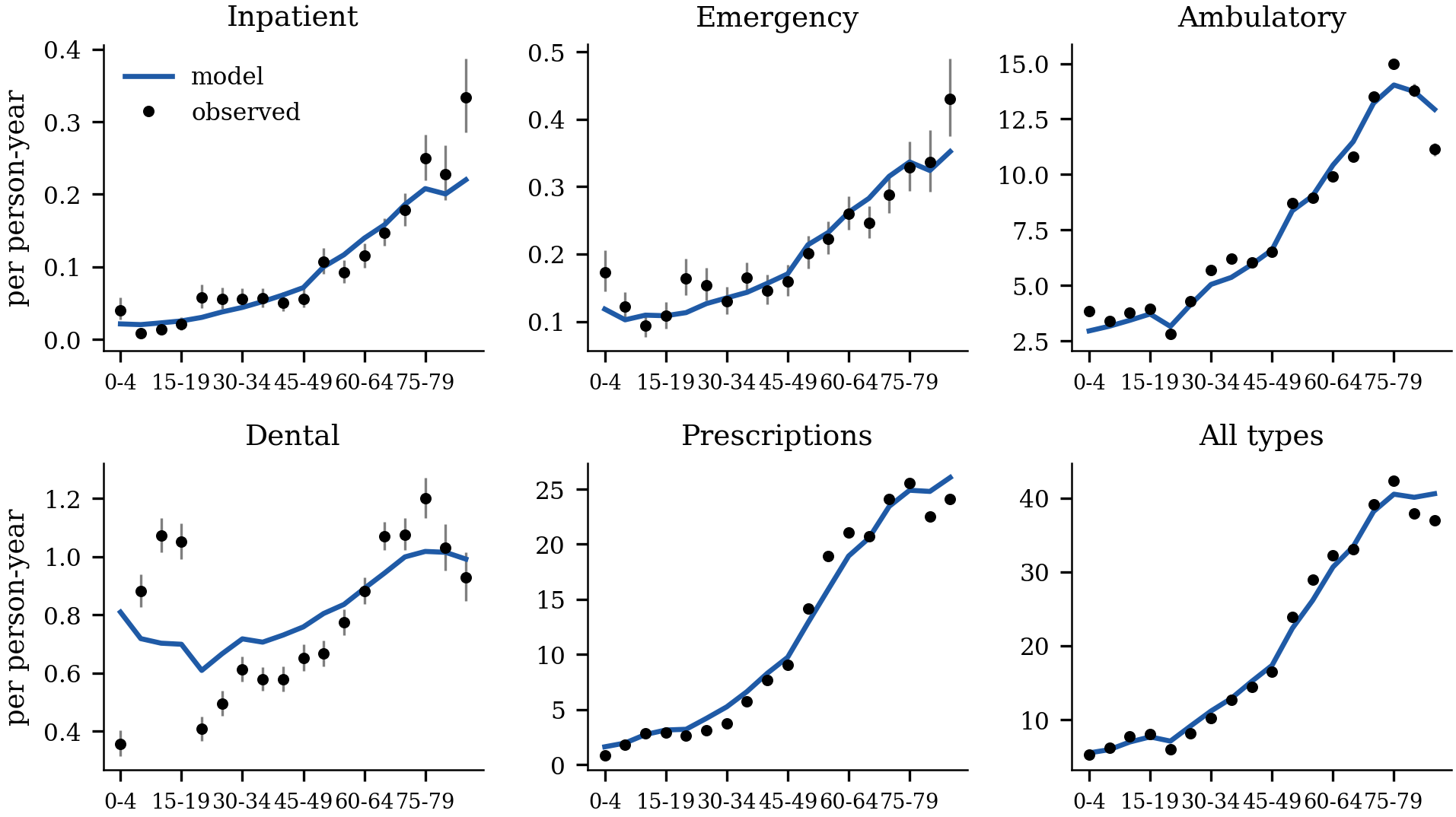}
\caption{Observed event frequency per person-year (points) and fitted rates (lines) by attained-age band. Scales differ between event types.}\label{fig:meps_age}
\end{figure}

For this application, we train the neural model for 100 epochs using a three-component mixture for conditional severity and independent type-specific gamma frailties. Optimization uses AdamW and a one-cycle learning-rate schedule.
Figure~\ref{fig:meps_age} compares observed event frequencies by attained age with those implied by the fitted model. The model closely follows the increase in ambulatory visits and prescriptions with age and also reproduces the higher rates of inpatient and emergency events among older individuals. The largest discrepancies occur for dental care, particularly among children and younger adults, and for inpatient events at the oldest ages. The fit is nonetheless good across much of the age range. 

\begin{figure}[!tbp]
\centering\includegraphics[width=0.98\linewidth]{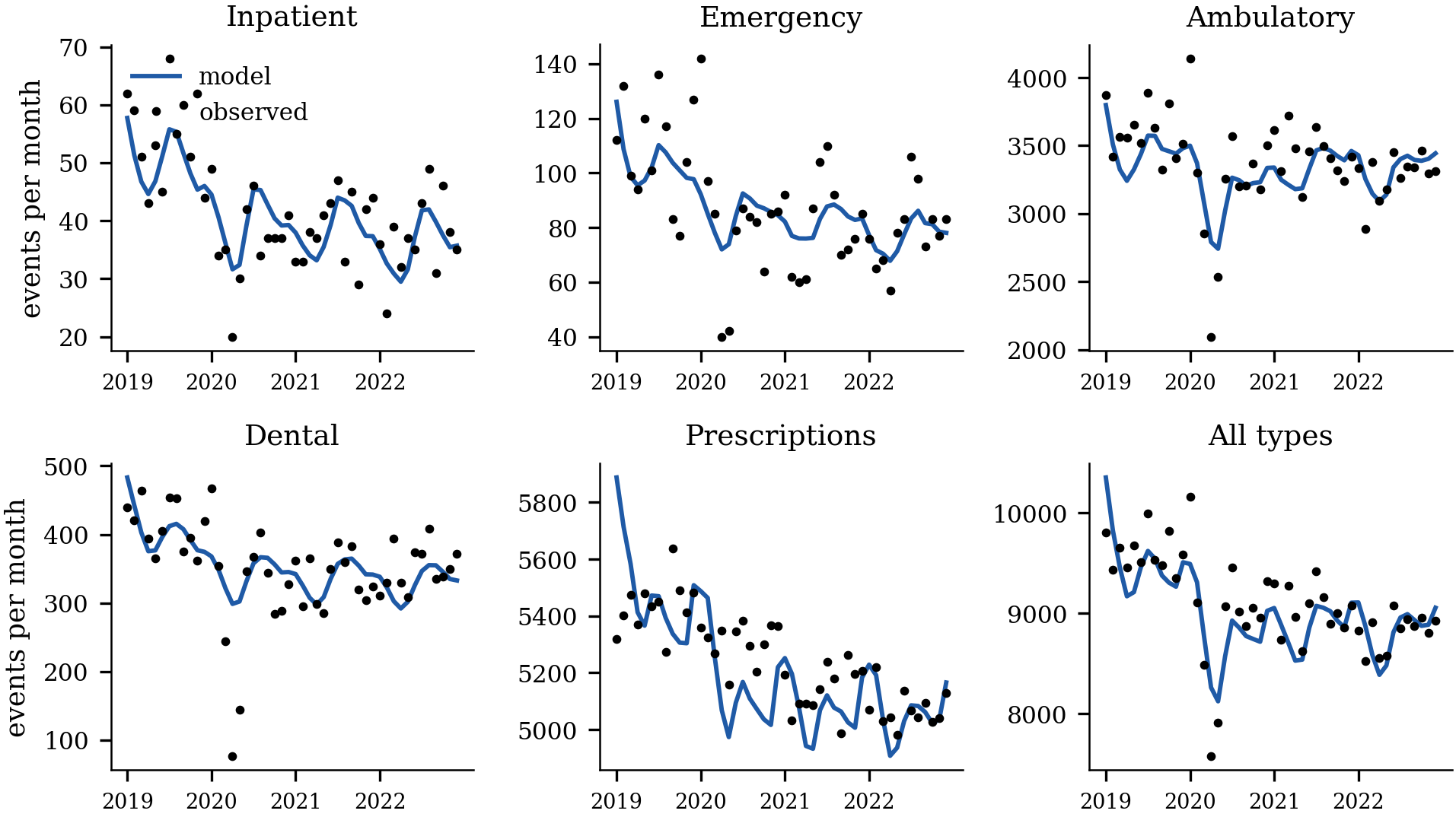}
\caption{Observed monthly event counts (points) and fitted counts (lines), separately by event type and for all events combined, 2019--2022.}\label{fig:meps_month}
\end{figure}

The monthly comparison in Figure~\ref{fig:meps_month} leads to a similar conclusion. The fitted series follow the main changes in utilization over 2019-2022, including the pronounced decline in several services during 2020. Agreement is especially close for the larger ambulatory and prescription series, although the model does not reproduce every monthly fluctuation in the less frequent event types. This comparison shows that the fitted intensities capture the main patterns across calendar time when evaluated using the recorded histories.

\begin{figure}[!tbp]
\centering\includegraphics[width=0.98\linewidth]{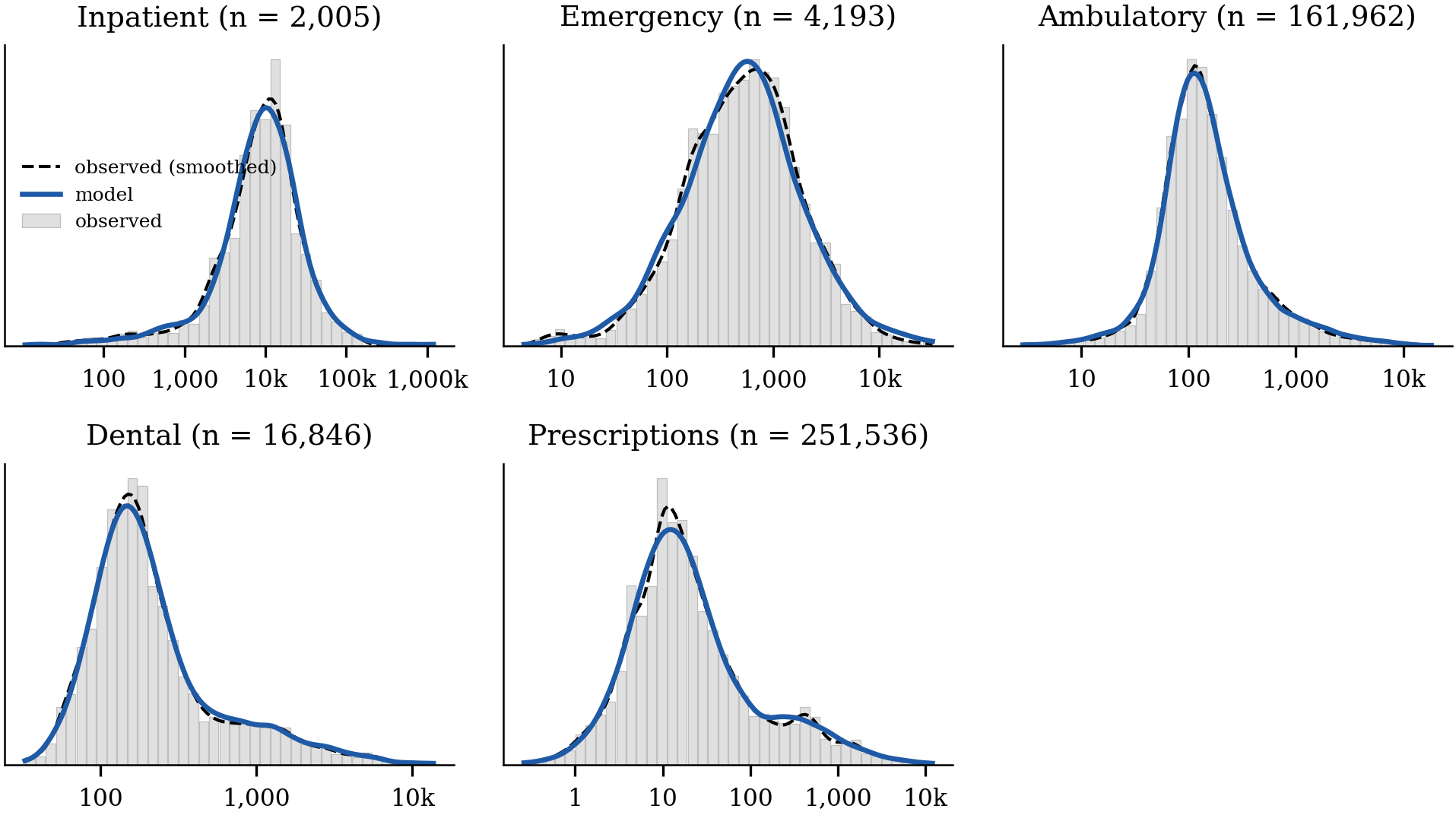}
\caption{Observed and fitted conditional event-amount distributions. Histograms and dashed curves describe the observations; solid curves show the fitted mixture densities. The horizontal axis is logarithmic.}\label{fig:meps_density}
\end{figure}

The preceding comparisons indicate that the fitted model reproduces the main age-related and calendar-time patterns in observed event frequencies when evaluated along the recorded claim histories. We now turn to the second component of the model, the conditional distribution of event expenditures. Figure~\ref{fig:meps_density} compares the observed expenditure distributions with those obtained from the fitted three-component lognormal mixtures. The model captures the main features of the distributions across all five event types, including the skewness and bimodality observed for dental services and prescriptions. Additional in-sample quantile and probability-integral-transform diagnostics are provided in Appendix~\ref{app:meps_diagnostics}. The predictive performance of the severity model is subsequently assessed using the 2022 holdout sample, which was not used to estimate the validation model.

\begin{figure}[!tbp]
\centering\includegraphics[width=0.98\linewidth]{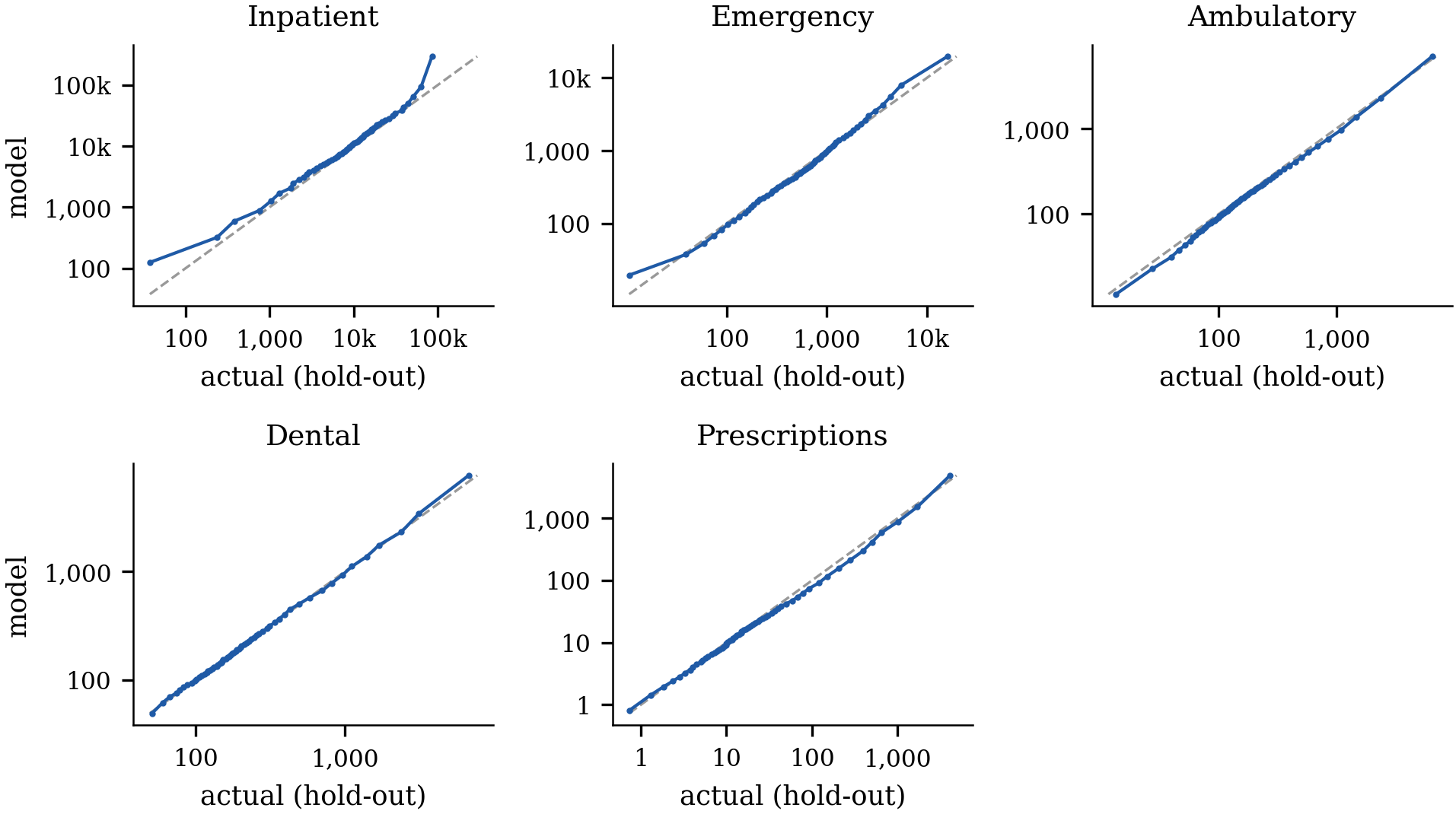}
\caption{Simulated versus observed claim-amount quantiles for events in the 2022 holdout, by type. The dashed line denotes equality.}\label{fig:meps_qq_oos}
\end{figure}

In the holdout, the model reproduces the distributions of ambulatory, dental and prescription amounts particularly well. Table~\ref{tab:meps_severity} shows that their mean amounts differ from the observed means by 1.3\%, 6.3\% and 0.7\%, respectively, while the corresponding 99th-percentile deviations are 4.7\%, 3.2\% and 1.7\%. Emergency severity is also reasonably close. Inpatient severity is less satisfactory: its simulated mean exceeds the observed mean by 56.6\% and its 99th percentile by 157.8\%. The median is also 24.1\% too high. This discrepancy is important because relatively few inpatient events account for a substantial share of expenditure. Figure~\ref{fig:meps_qq_oos} confirms that the main departures from the observed severity quantiles occur for the largest inpatient amounts.

\begin{table}[!tbp]
\centering\small
\caption{Conditional severity in the 2022 holdout. Relative deviations are simulated divided by observed, minus one. The simulated distribution is evaluated using the history available before each observed event.}\label{tab:meps_severity}
\setlength{\tabcolsep}{5pt}
\begin{tabular}{lrrrr}
\toprule
Type & Events & Mean deviation & Median deviation & 99th quantile deviation\\
\midrule
Inpatient & 445 & +56.6\% & +24.1\% & +157.8\%\\
Emergency & 947 & +5.5\% & -8.4\% & +9.8\%\\
Ambulatory & 39,351 & +1.3\% & -3.2\% & +4.7\%\\
Dental & 4,220 & +6.3\% & +1.8\% & +3.2\%\\
Prescriptions & 60,881 & +0.7\% & +3.5\% & +1.7\%\\
\bottomrule
\end{tabular}
\end{table}

\begin{figure}[!tbp]
\centering\includegraphics[width=0.95\linewidth]{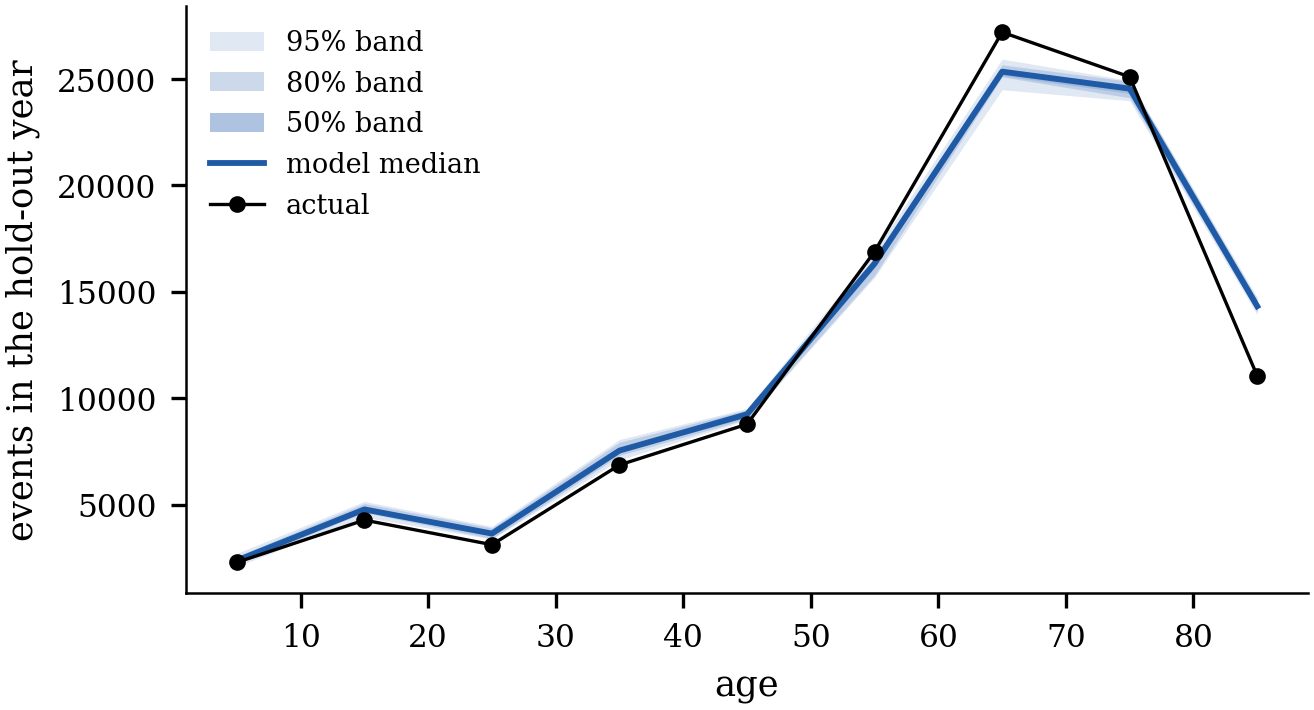}
\caption{Total events in the 2022 holdout by age group. The line shows the simulated median, shading shows central 50\%, 80\% and 95\% ranges, and points show observed counts.}\label{fig:meps_holdout_age}
\end{figure}

We assess forecasts for 2022 by fitting the validation model on 2019-2021 and simulating the following year for the individuals in the test population. For the common set of 5,147 individuals, the observed total is 105,546 events, compared with a mean of 102,310 across the simulated paths. The corresponding actual-to-predicted ratio is 1.032, close to the ratio of 1.031 from the annual benchmark on the same population. Figure~\ref{fig:meps_holdout_age} provides a more detailed comparison by age. The simulated counts follow the observed distribution closely through most age groups, including the marked increase in utilization from middle to older ages. The largest visible discrepancy occurs in the oldest group, where the model overpredicts the number of events. Thus, the modest difference in annual totals is accompanied by a broadly accurate age pattern, although the agreement is not uniform across the population.

\begin{figure}[!tbp]
\centering\includegraphics[width=0.98\linewidth]{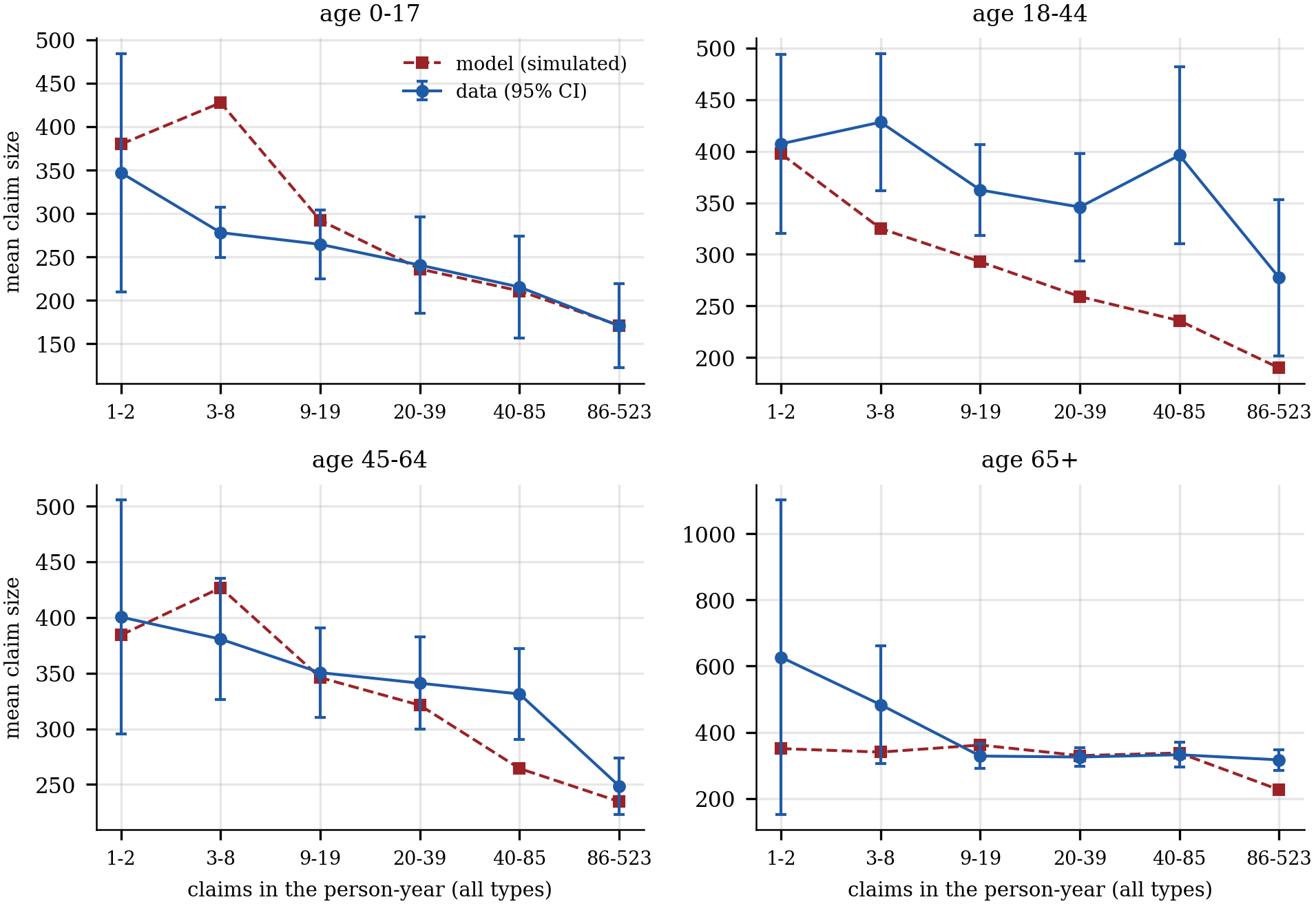}
\caption{Mean event amount by annual event-count category and age group. Solid points and vertical bars show the observed mean and its displayed 95\% confidence interval; dashed lines show means from the model simulations. Count categories refer to the number of events of all types in a person-year.}\label{fig:meps_frequency_severity}
\end{figure}

The relationship between the number of events and their average expenditure is examined in Figure~\ref{fig:meps_frequency_severity}. Individuals are separated into four age groups and into categories according to the number of events in a person-year, with the model-simulated mean claim amount compared with the observed mean and its displayed 95\% confidence interval. Across the age groups, average amounts tend to be lower in person-years with more events, although the pattern is not identical at every age. The model captures much of this decline, particularly among children and individuals aged 45--64. For ages 18--44, however, simulated mean amounts are generally below the observed values once annual counts exceed a few events; among those aged 45--64, the discrepancy is most apparent at higher counts. The oldest age group shows broadly similar means in the intermediate count categories, but the comparison is less precise where relatively few observations are available. 
This relationship is particularly relevant because
conventional actuarial models often treat claim
frequency and severity as independent. In contrast,
the proposed model allows both claim occurrence
and claim amounts to depend on the recorded
event history, providing a mechanism for
capturing this dependence.

\begin{figure}[!tbp]
\centering\includegraphics[width=0.98\linewidth]{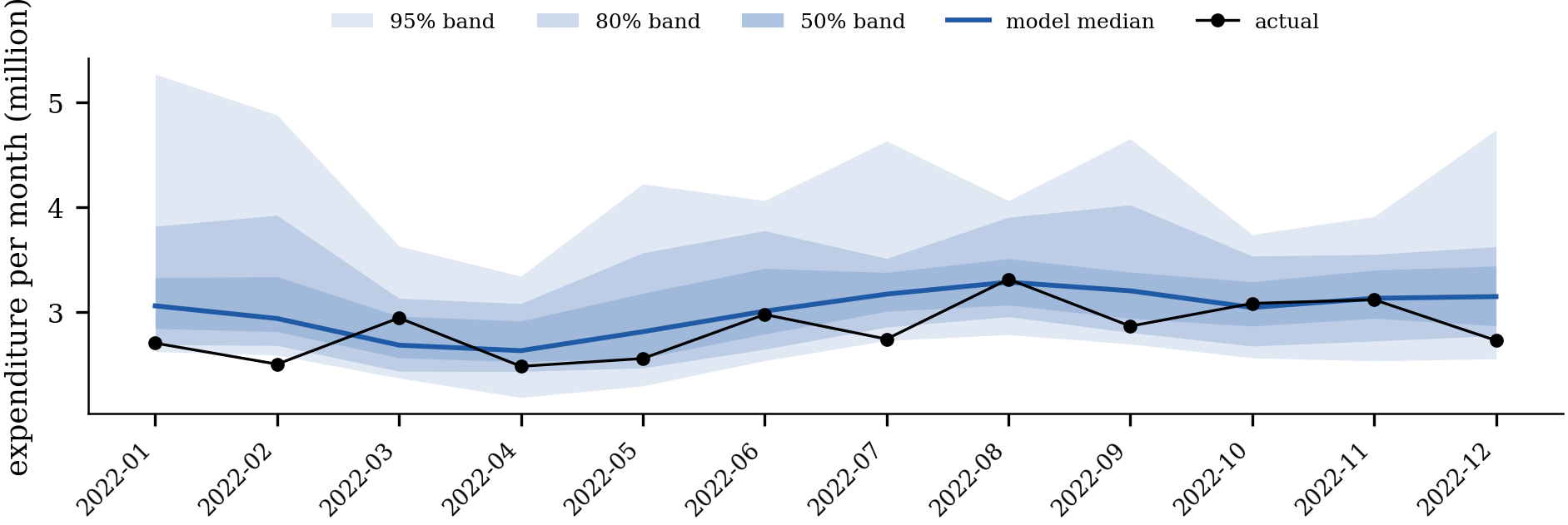}
\caption{Monthly aggregate expenditure in the 2022 holdout. The solid line shows the simulated median, shaded areas central 50\%, 80\% and 95\% ranges, and points the observed amounts.}\label{fig:meps_holdout_cost_month}
\end{figure}

The monthly expenditure forecasts in Figure~\ref{fig:meps_holdout_cost_month} are generally of the same order as the observed totals, with appreciable variation among simulated paths. This agreement at the portfolio level does not remove the substantial error in the upper tail of inpatient severity. In particular, predictions of total expenditure can be reasonable even when one costly event category is less accurately modeled, since the more frequent categories contribute much of the overall expenditure.

Finally, we compare the neural model with two annual generalized linear models. GLM-0 uses demographic and calendar-year characteristics, while GLM-H also includes summaries of utilization in the preceding year. The comparison uses the same 2022 holdout population. As Table~\ref{tab:meps_benchmark} shows, the neural model achieves the lowest pooled Poisson count deviance, at 8.24 compared with 10.66 for GLM-H and 24.77 for GLM-0. It also has the highest expenditure-ranking Gini, at 0.752 compared with 0.678 and 0.432, and its highest-ranked decile contains 40.3\% of observed expenditure. This is a material improvement in the overall ranking of expenditure risk, including relative to a GLM that already uses past utilization. 

\begin{table}[!tbp]
\centering\small
\caption{Annual 2022 holdout benchmark. Count deviance is mean Poisson deviance (lower is better); higher Gini values and top-decile shares indicate stronger ranking. A/E is observed divided by predicted counts or expenditure.}\label{tab:meps_benchmark}
\setlength{\tabcolsep}{5pt}
\begin{tabular}{lrrrrrr}
\toprule
Model & Count deviance & Count Gini & Cost Gini & Top 10\% & Count A/E & Cost A/E\\
\midrule
GLM-0 & 24.77 & 0.535 & 0.432 & 17.6\% & 0.930 & 0.350\\
GLM-H & 10.66 & 0.874 & 0.678 & 34.2\% & 1.126 & 0.113\\
Neural & 8.24 & 0.892 & 0.752 & 40.3\% & 1.031 & 0.962\\
\bottomrule
\end{tabular}
\end{table}

\begin{figure}[!tbp]
\centering\includegraphics[width=0.98\linewidth]{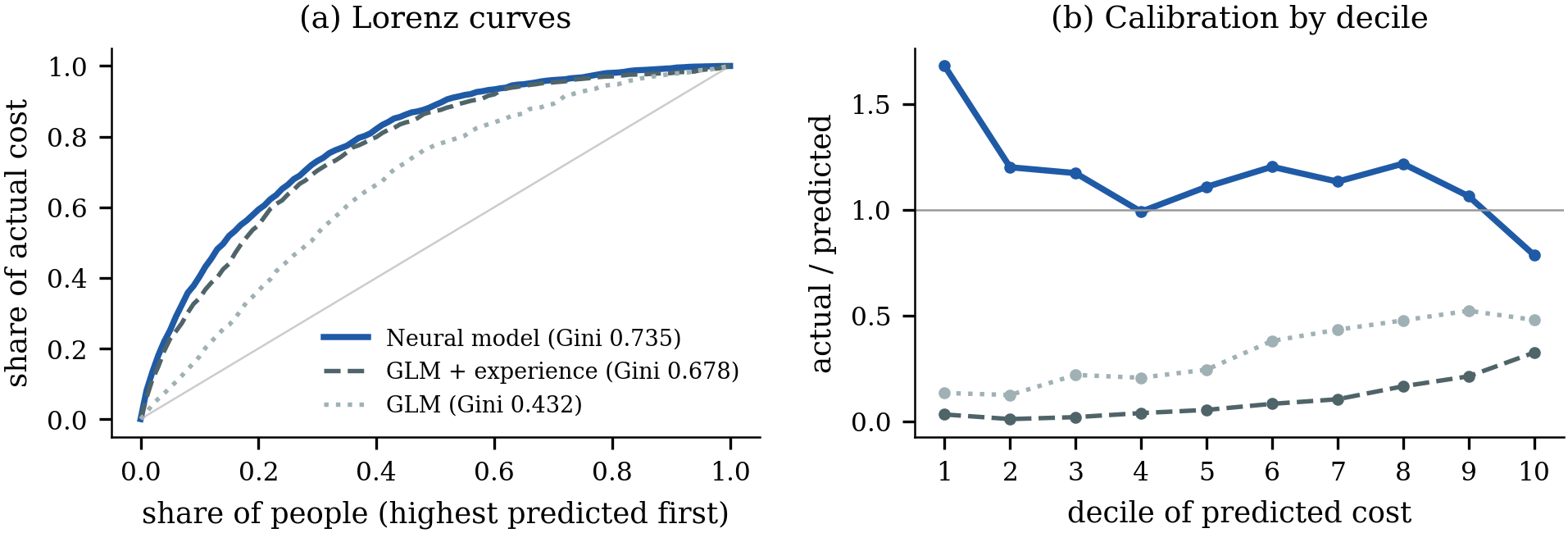}
\caption{Expenditure ranking (left) and observed-to-predicted expenditure by decile of predicted cost (right) for the annual 2022 holdout.}\label{fig:meps_benchmark}
\end{figure}

Ranking and calibration are different properties. The neural model's pooled actual-to-predicted expenditure ratio is 0.962, whereas the corresponding ratios for GLM-0 and GLM-H are 0.350 and 0.113. Thus the GLMs rank individuals less well and their reported monetary predictions are also much higher than the observed expenditure. 

The MEPS results give a clear picture of both the strengths and the remaining limitations of the approach. The fitted model describes age-specific utilization, calendar-time frequencies and most conditional expenditure distributions well. The frequency--severity comparison also shows that it captures much of the decrease in average claim amounts as annual utilization increases, although the fit varies by age. It ranks future expenditure risk more effectively than the two GLM benchmarks. At the same time inpatient severity is overestimated, especially for large amounts. 

\section{Discussion and conclusion}\label{sec:discussion}
This paper develops a neural marked point-process
framework for modeling individual health insurance
claims. By combining a Transformer representation
of previous claim histories with type-specific
conditional intensities and a flexible distribution
for claim amounts, the model allows both the
frequency and severity of claims to depend on
the sequence of earlier events. Individual
heterogeneity is incorporated through gamma
frailties, while medical inflation and reporting
delays are accounted for within the calibration
framework. The resulting approach provides
an alternative to models based on aggregate
claim counts and amounts, particularly when
the timing and ordering of individual events
contain information about subsequent claim
experience.

We evaluate the approach using a controlled
simulation and an empirical application to
medical-event histories from the Medical
Expenditure Panel Survey. The simulation
shows that the model captures several
important features of a data-generating
process that differs from the fitted neural
specification, including age-related
differences in claim frequency, variation
in claim amounts and dependence between
successive events. The empirical application
provides further evidence of its predictive
capabilities. The model reproduces the
main patterns in healthcare utilization
and expenditure and achieves lower pooled
count deviance and better expenditure
ranking than the two generalized linear
models considered, including a specification
that incorporates historical utilization.

An important finding of the empirical analysis is
that the model captures not only differences in
average claim frequencies and expenditures, but
also several features of individual healthcare
utilization. The results illustrate the potential
of combining flexible history representations
with models for claim frequency and severity.
At the same time, the comparison between fitted
and simulated event histories provides insights
into the role of persistent individual
heterogeneity and its interaction with previous
claims. This highlights the value of examining
the evolution of individual claim histories
alongside conventional measures of predictive
accuracy.

The proposed framework also offers several
directions for further research. More flexible
representations of individual heterogeneity
could improve the description of long-term
utilization patterns, while comparisons with
alternative neural and statistical models
would provide further insight into the benefits
of explicitly incorporating claim histories.
Applications to real insurance data
with detailed occurrence and reporting
information would be particularly valuable,
allowing the approach to be examined in a
setting that more closely reflects practical
insurance operations. Such applications
could also extend the empirical evaluation
to reporting delays and claims reserving.

Overall, the findings support the use
of neural marked point processes as a
flexible approach to modeling individual
medical-event histories. Their main
advantage lies in the ability to
incorporate the information contained
in previous events into predictions
of subsequent claim occurrences and
amounts, while retaining a probabilistic
description of the underlying processes.

\section*{Acknowledgements}
The authors would like to thank M. Coriandre for the invaluable
discussions and for providing the inspiration for
this research.

The authors acknowledge the Agency for Healthcare
Research and Quality (AHRQ) for making the Medical
Expenditure Panel Survey (MEPS) public-use data
available. The analyses and conclusions presented
in this paper are those of the authors and do not
necessarily reflect the views of AHRQ or the
U.S. Department of Health and Human Services.

\section*{Declaration of competing interests}

The authors declare that they have no known
financial or non-financial competing interests
that could have influenced the research reported
in this paper.

\section*{Declaration of generative AI use}

During the preparation of this manuscript, the authors used ChatGPT (OpenAI, GPT-6) to refine the text and formulas, and support the preparation of figures. Claude (Anthropic) was also used to assist with Python coding for the numerical experiments. The authors reviewed and adapted the AI-assisted text and mathematical expressions and take full responsibility for the final manuscript, including its mathematical results, computational methods, numerical findings, and references.
\section*{Funding}
This research was funded within the Post-Doc Track program of the Austrian Academy of Sciences.
\appendix

\section{Additional MEPS calibration diagnostics}\label{app:meps_diagnostics}

\begin{figure}[!tbp]
\centering\includegraphics[width=0.98\linewidth]{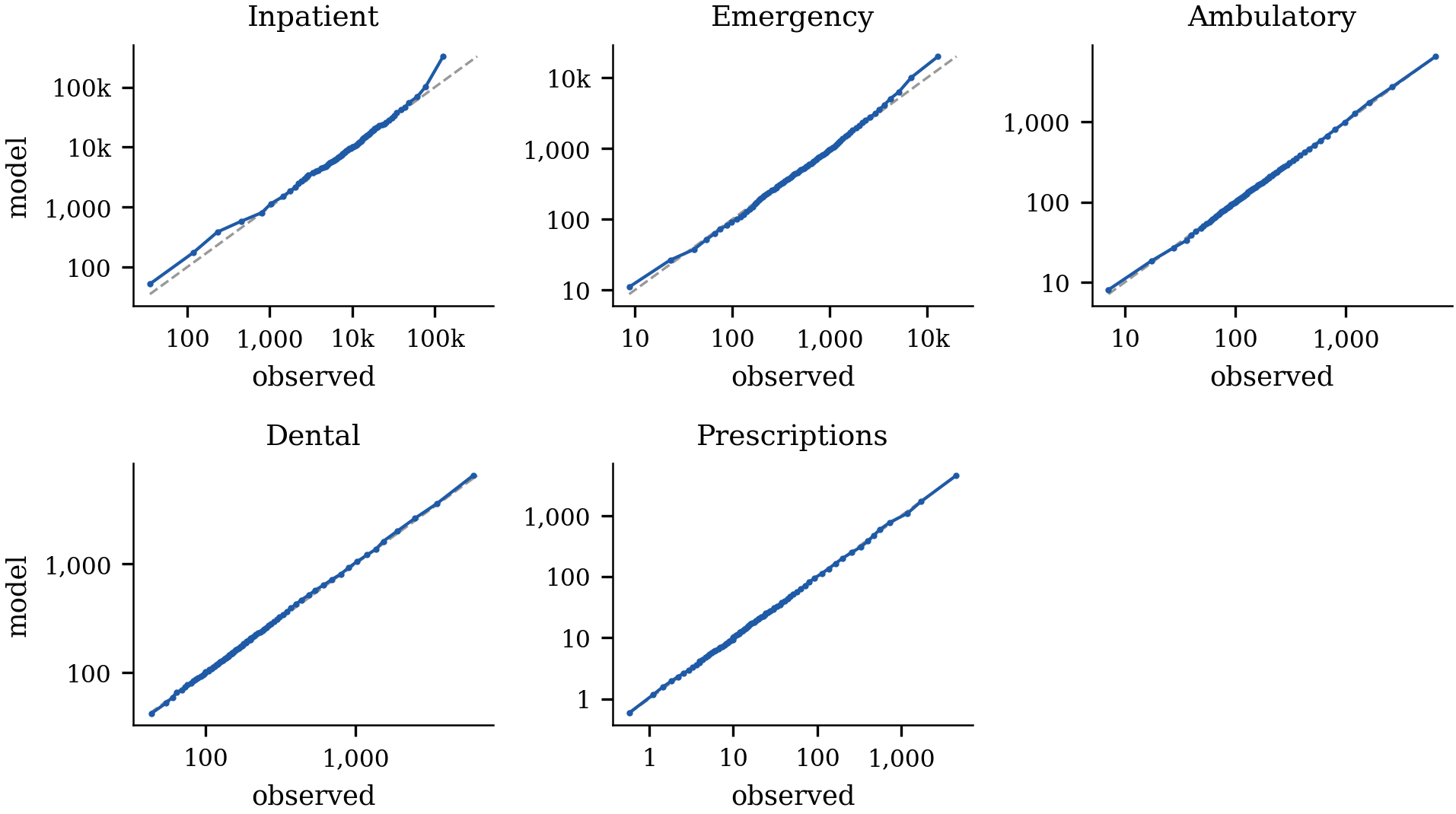}
\caption{In-sample severity QQ plots, processed MEPS events. The diagonal indicates quantile agreement on the displayed logarithmic scales.}\label{fig:meps_qq_insample}
\end{figure}

\begin{figure}[!tbp]
\centering\includegraphics[width=0.98\linewidth]{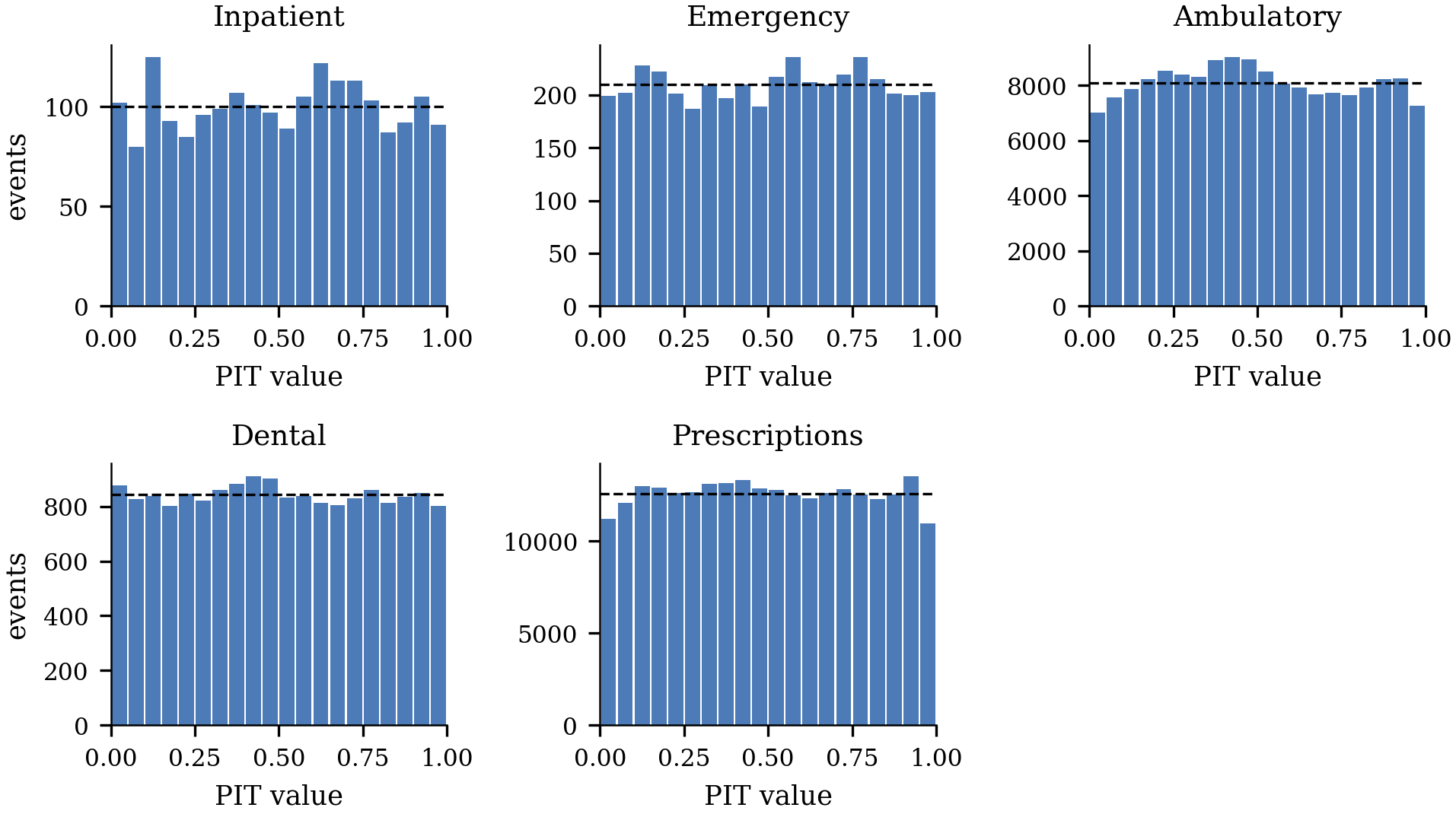}
\caption{In-sample probability-integral-transform histograms for event amounts, by service type. Dashed lines denote the uniform-reference bin counts; these plots are descriptive in-sample checks.}\label{fig:meps_pit}
\end{figure}

Figure~\ref{fig:meps_qq_insample} gives the within-sample severity quantile comparison and Figure~\ref{fig:meps_pit} the probability-integral-transform histograms for recorded event amounts. These checks complement the held-out QQ plot in Figure~\ref{fig:meps_qq_oos}. Small departures from uniformity are visible, but the PIT diagrams are based on fitted observations and are not an independent test of forecasting performance. The full supplied age-by-type, severity, holdout and benchmark tables are preserved as supplemental source files.

\end{document}